\UseRawInputEncoding
\documentclass[
 aip,
 amsmath,amssymb,
reprint,%
]{revtex4-1}

\newif\ifarxiv
\arxivtrue

\usepackage{color}
\usepackage{here}
\usepackage{graphicx}
\usepackage{dcolumn}
\usepackage{bm}
\usepackage{cases}

\ifarxiv
\else
\usepackage{lineno}
\newcommand*\patchAmsMathEnvironmentForLineno[1]{%
  \expandafter\let\csname old#1\expandafter\endcsname\csname #1\endcsname
  \expandafter\let\csname oldend#1\expandafter\endcsname\csname end#1\endcsname
  \renewenvironment{#1}%
     {\linenomath\csname old#1\endcsname}%
     {\csname oldend#1\endcsname\endlinenomath}}%
\newcommand*\patchBothAmsMathEnvironmentsForLineno[1]{%
  \patchAmsMathEnvironmentForLineno{#1}%
  \patchAmsMathEnvironmentForLineno{#1*}}%
\AtBeginDocument{%
\patchBothAmsMathEnvironmentsForLineno{equation}%
\patchBothAmsMathEnvironmentsForLineno{align}%
\patchBothAmsMathEnvironmentsForLineno{flalign}%
\patchBothAmsMathEnvironmentsForLineno{alignat}%
\patchBothAmsMathEnvironmentsForLineno{gather}%
\patchBothAmsMathEnvironmentsForLineno{multline}%
}
\linenumbers
\fi

\usepackage[T1]{fontenc}
\usepackage{newpxtext, mathptmx}

\newcommand{\Alfven}{Alfv\'en }
\newcommand{\para}{\parallel}
\newcommand{\Wci}{\Omega_{i}}
\newcommand{\Wcr}{\Omega_{r}}
\newcommand{\Wce}{\Omega_{e}}
\newcommand{\wpi}{\omega_{pi}}
\newcommand{\wpr}{\omega_{pr}}
\newcommand{\wpe}{\omega_{pe}}
\newcommand{\vth}[1]{v_{{\rm th}, {#1}}}
\newcommand{\VA}{V_{\rm A}}
\newcommand{\MA}{M_{\rm A}}
\newcommand{\MS}{M_{\rm s}}

\begin{document}

\preprint{AIP/123-QED}

\title[Ion-scale Kinetic Instability at Shock]{Mach Number Dependence of Ion-scale Kinetic Instability at Collisionless Perpendicular Shock: Condition for Weibel-dominated Shock}

\author{Takuro Nishigai}
\author{Takanobu Amano}%
\email{amano@eps.s.u-tokyo.ac.jp}
\affiliation{
Department of Earth and Planetary Science, The University of Tokyo, \\
7-3-1 Hongo, Bunkyo-ku, Tokyo 113-0033, Japan.
}%

\date{\today}

\begin{abstract}
We investigate ion-scale kinetic plasma instabilities at the collisionless shock using linear theory and nonlinear Particle-in-Cell (PIC) simulations. We focus on the Alfv\'en-ion-cyclotron (AIC), mirror, and Weibel instabilities, which are all driven unstable by the effective temperature anisotropy induced by the shock-reflected ions within the transition layer of a strictly perpendicular shock. We conduct linear dispersion analysis with a homogeneous plasma model to mimic the shock transition layer by adopting a ring distribution with finite thermal spread to represent the velocity distribution of the reflected ions. We find that, for wave propagation parallel to the ambient magnetic field, the AIC instability at lower \Alfven Mach numbers tends to transition to the Weibel instability at higher \Alfven Mach numbers. The instability property is, however, also strongly affected by the sound Mach number. We conclude that the instability at a strong shock with \Alfven and sound Mach numbers both in excess of $\sim 20{\rm -}40$ may be considered as Weibel-like in the sense that the reflected ions behave essentially unmagnetized. Two-dimensional PIC simulations confirm the linear theory and find that, with typical parameters of young supernova remnant shocks, the ring distribution model produces magnetic fluctuations of the order of the background magnetic field, which is smaller than those observed in previous PIC simulations for Weibel-dominated shocks. This indicates that the assumption of the gyrotropic reflected ion distribution may not be adequate to quantitatively predict nonlinear behaviors of the dynamics in high Mach number shocks.
\end{abstract}

\maketitle

\section{Introduction}
\label{sec:intro}
The dynamics of shock waves in collisionless plasmas have been a subject of extensive research over the past few decades. A hydrodynamic shock wave must, by definition, involve the dissipation of the supersonic upstream flow kinetic energy via viscosity at the shock front. The dissipation provides the increased thermal pressure in the decelerated downstream plasma to sustain the stable shock structure. Since the viscosity in the classical sense is negligible in a collisionless plasma, it should be provided effectively by complicated nonlinear interactions between electromagnetic fluctuations and charged particles. Understanding the origin of fluctuations or waves in and around the collisionless shocks is thus of critical importance.

It has long been recognized that the collisionless shock is a "zoo" of plasma kinetic instabilities \citep{Wu1984a}. In a magnetized plasma, the transverse component of the magnetic field is compressed at a thin shock transition layer. The current associated with the magnetic field compression across the shock may trigger microscopic instabilities. Scattering of particles by enhanced electromagnetic fluctuations may contribute to the dissipation via an anomalous resistivity. However, it is well known that, if the \Alfven Mach number exceeds a threshold,  the resistive heating alone is not able to suppress the steepening of the shock to keep the structure stable \citep{Coroniti1970}. The threshold, known as the critical Mach number, is theoretically determined to be $\sim 2.7$ at most \citep{Kennel1985}, indicating most of shock waves seen in space plasmas are super-critical shocks. In-situ observations and numerical simulations of super-critical shocks clearly indicate that upstream ions coming into the shock are partially reflected at the shock \citep{Leroy1981,Leroy1982,Paschmann1982,Sckopke1983}. The reflected ions gyrating around the upstream magnetic field gain energies by traveling along the upstream motional electric field, which can provide a substantial contribution to the downstream pressure. At the same time, they drive various kinds of beam-plasma instabilities as they appear as a distinct beam population in velocity space on top of the upstream thermal distribution. In this way, the reflected ions play the dominant role in the dynamics of the collisionless shocks.

The characteristics and consequences of instabilities driven by the reflected ions depend on the parameters of the shock. One of the most important parameters controlling the dynamics of the shock is the magnetic field orientation with respect to the normal of the shock surface. To simplify the discussion, in this paper, we will concentrate on the strictly perpendicular shock, in which the magnetic field makes an angle of $\theta_{Bn} = 90^{\circ}$ to the shock normal. However, the result will also be applied to quasi-perpendicular shocks with $\theta_{Bn} > 45^{\circ}$ because the reflected-ion dynamics is very similar in these shocks. In addition, the dynamics of quasi-perpendicular shock is not entirely unrelated to the quasi-parallel shock ($\theta_{Bn} < 45^{\circ}$), as the transition layer of a quasi-parallel shock might appear locally like a quasi-perpendicular shock because of amplification of the transverse magnetic field occurring in the immediate upstream of the shock.

It is easy to conjecture that the Mach number is another important parameter for the kinetic instabilities at the collisionless shocks. In a very high Mach number regime, the relative velocity between the beam and the upstream plasma may exceed the electron thermal velocity, in which violent electrostatic Buneman instability is triggered \citep{Cargill1988,PapadopoulosK.1988,Shimada2000}. Strong electron heating and acceleration that may result from large-amplitude electrostatic waves driven by the Buneman instability have been investigated using fully kinetic Particle-In-Cell (PIC) simulations \citep{Dieckmann2000a,McClements2001,Hoshino2002,Dieckmann2004a,Amano2007,Matsumoto2013,Bohdan2019a}. At lower Mach numbers, such high-frequency electrostatic waves will be stabilized by the electron Landau damping. Instead, the modified two-stream instability may generate electromagnetic waves on the oblique whistler-mode branch at wavelengths comparable to the electron inertial length. Efficient heating of both ions and electrons associated with oblique whistlers have been discussed in the literature \citep{Scholer2003a,Matsukiyo2003b,Matsukiyo2006b,Umeda2012a,Umeda2012b}.

At the ion scale, an effective temperature anisotropy ($T_{\perp}/T_{\para} > 1$ with $T_{\perp}$ and $T_{\para}$ being perpendicular and parallel temperatures) due to the reflected ions gyrating around a local magnetic field may become the source of instability. The so-called rippling mode identified in early numerical simulations has been believed to be the consequence of either Alf\'ven-ion-cyclotron (AIC) or mirror instabilities \citep{Winske1988}. Oscillatory features observed in the terrestrial bow shock are sometimes attributed to the rippling mode \citep{Johlander2016,Johlander2018}. More recent PIC simulations for high Mach number non-relativistic perpendicular shocks, relevant to young supernova remnant (SNR) shock, suggested that the electromagnetic fluctuations at the ion scale driven by the reflected ions are instead generated by the Weibel instability \citep{Kato2010,Matsumoto2015,Matsumoto2017,Bohdan2020a,Bohdan2021}. In contrast to the rippling mode, in which fluctuating magnetic field amplitudes are of the same order of the background field, the Weibel instability amplifies the magnetic field to a level much larger than the background field. Folded current sheets generated as a result of the magnetic field amplification are then spontaneously dissipated via magnetic reconnection. The interesting finding of magnetic reconnection within the shock transition layer suggests that there may be completely different mechanisms than previously thought for the heating and acceleration of particles at the shock \citep{Matsumoto2015,Bohdan2020a}. It is, however, not very well understood at which condition the dominant instability transitions from the AIC/mirror (or rippling) to the more violent Weibel instability.

It should be mentioned that the importance of the Weibel instability in the context of collisionless shocks has been widely recognized for a highly relativistic unmagnetized plasma \citep{Medvedev1999b,Silva2003,Medvedev2005,Kato2005,Achterberg2007a,Achterberg2007b,Kato2008,Spitkovsky2008,Bret2013}. Since effective particle heating is provided by scattering via Weibel-generated turbulence, such a shock is often termed as a Weibel-mediated shock. On the other hand, we are interested in non-relativistic magnetized shocks, in which the shock structure is controlled mainly by reflected-gyrating ions in the background magnetic field. In this paper, a non-relativistic magnetized shock that is substantially affected by the dominant Weibel instability will be referred to as a Weibel-dominated shock. It is no surprise that the Weibel-dominated shocks substantially affect particle heating and acceleration mechanisms suggested previously based on the classical magnetized shock structure. Therefore, understanding the transition from the classical to Weibel-dominated shock provides crucial information also for estimating the heating and acceleration efficiencies at high Mach number collisionless shocks.

The purpose of this paper is to investigate the relationship between AIC/mirror and Weibel instabilities, all driven by the same equilibrium condition, in collisionless perpendicular shocks. In particular, we focus our attention on the parameter dependence of the instability characteristics. We should mention that the AIC, mirror, and Weibel instabilities have been extensively investigated both theoretically and numerically. The AIC and mirror instabilities assume that the plasma is magnetized, whereas the Weibel instability usually considers an unmagnetized plasma as the equilibrium state. In the context of collisionless shocks, the plasma should be treated as magnetized if the upstream magnetic field is finite. As the \Alfven Mach number increases, the growth rate of an instability relative to the gyrofrequency may become larger and larger because of stronger instability driving. At some point, the growth rate may become larger than the gyrofrequency, in which case, the unmagnetized plasma approximation should become reasonable. In this way, we may conjecture that either AIC or mirror instabilities (or possibly a mixture of them) for a magnetized plasma at lower \Alfven Mach number continuously changes its property to the Weibel instability in a nearly unmagnetized plasma at sufficiently high \Alfven Mach numbers. In this paper, we will demonstrate that this is indeed the case. For this purpose, we employ a magnetized ring distribution for the reflected ion component, which allows us to study the continuous transition of instability characteristics in a wide range of the shock parameters.

This paper is organized as follows. In Section \ref{sec:model}, we first introduce the model that we use throughout in this study. The result of linear dispersion analysis will be shown in Section \ref{sec:linear}, in which we demonstrate a continuous transition of instability characteristics between the two different regimes. In Section \ref{sec:simulation}, two-dimensional (2D) PIC simulation results will be discussed. We have confirmed the linear theory and also found that the saturation of the instability in the Weibel regime is controlled by a particle trapping criterion. Finally, discussion and conclusions will be given in Section \ref{sec:discussion}

\section{Model}
\label{sec:model}
It is obvious that the structure of collisionless shock involves strong spatial inhomogeneity for both electric and magnetic fields. The inhomogeneity is indeed crucial for the ion reflection. We will not, however, take into account the self-consistent ion reflection in this paper. Instead, we consider a spatially homogeneous system in which the reflected ions exist in the first place. The periodic model has often been employed to simplify the problem as much as possible \citep{Matsukiyo2003b,Matsukiyo2006b,Muschietti2013}. It must be mentioned that the model is not designed for an exact match with what happens in a fully self-consistent collisionless shock structure. It should rather be recognized as a simplified model to extract the essential physics.

In most of the previous works employing a periodic model, the reflected ions are approximated as a distinct non-gyrotropic (or unmagnetized) beam population. Such an assumption should be valid for waves with wavelengths much less than the ion gyroradius or, equivalently, with frequencies much higher than the ion gyrofrequency. Although we expect that the Weibel   instability should satisfy the condition, the AIC and mirror instabilities do not. Since our purpose is to obtain a unified understanding of the ion-scale kinetic instabilities both for the unmagnetized and magnetized regimes, we should consider something different.

In this paper, we consider a homogeneous plasma in the background magnetic field $B_0$, consisting of three particle populations; the background electrons, background ions, and reflected ions. We assume that the reflected ions constitute a ring distribution with finite thermal spread with respect to the center of the ring in velocity space \citep{McClements1993,McClements1997}. With this model, we intend to approximate the gyrating particle distribution immediately after the reflection at the shock front. One may consider such a distribution gives a reasonable approximation for the velocity distribution function integrated over the transition layer of a perpendicular shock. We use the following definition for the ring distribution (with subscript $r$ indicating the reflected ions):
\begin{align}
    F_{r} (v_{\perp}, v_{\para}) = F_0
    \exp
    \left( 
        - \frac{v_{\para}^2}{2 \vth{r}^2}
        - \frac{(v_{\perp} - V_{r})^2}{2 \vth{r}^2}
    \right),
    \label{eq:ring}
\end{align}
where
\begin{align}
    \frac{1}{F_0} &= \left( 2 \pi \vth{r}^2 \right)^{3/2}
    \times \\
    &
    \left[
        \exp \left( - \frac{V_{r}^2}{2 \vth{r}^2} \right) +
        \sqrt{\pi} \frac{V_{r}}{\sqrt{2} \vth{r}}
        {\rm erfc} \left( - \frac{V_{r}}{\sqrt{2} \vth{r}} \right)
    \right]
\end{align}
is a constant normalization factor. We assume the zeroth order velocity distribution is gyrotropic and can be written as a function of parallel ($v_{\para}$) and perpendicular ($v_{\perp}$) velocities with respect to the magnetic field direction. Henceforth, $V_{r}$ will be referred to as the ring velocity. Note that ${\rm erfc}(x)$ represents the complementary error function of argument $x$. The background electrons ($s = e$) and ions ($s = i$) are assumed to be an isotropic Maxwellian:
\begin{align}
    F_{s} (v_{\perp}, v_{\para}) =
    \frac{1}{\left( 2 \pi \vth{s}^2 \right)^{3/2}}
    \exp \left(
        - \frac{v_{\para}^2}{2 \vth{s}^2} - \frac{v_{\perp}^2}{2 \vth{s}^2}
    \right),
    \label{eq:maxwellian}
\end{align}
We call $\vth{s}$ as the thermal velocity for species $s$, although it has a slightly different meaning for the perpendicular spread of the ring distribution. Note that the distribution function Eqs.~(\ref{eq:ring}) and (\ref{eq:maxwellian}) are normalized to give a unit density and should be multiplied by the density $n_{s}$. We define the background density $n_0$ to be equal to the electron density $n_0 = n_e = n_r + n_i$, where the last equality is required by the charge neutrality condition.

In the following, we conduct theoretical and numerical analyses in the reference frame in which all the three populations are at rest. It should be mentioned that the bulk velocity of the ring distribution given by Eq.~(\ref{eq:ring}) is zero because of symmetry in the plane perpendicular to the magnetic field. Instead, a finite ring velocity $V_{r} \neq 0$ introduces an effective perpendicular temperature $T_{\perp} \propto V_{r}^2$ (for a cold ring distribution with $V_{r} \gg \vth{r}$), which gives the free energy for instability. The units for time and velocity are, respectively, the inverse ion gyrofrequency $1/\Wci = m_i c / e B_0$ and the Alfven speed $\VA = B_0/\sqrt{4 \pi n_0 m_i}$. The corresponding length scale $\VA/\Wci$ defines the ion inertial length. Note that we use the standard notations: $c$ for the speed of light, $e$ for the elementary charge, $m_{s}$ for the mass of species $s$.

Let us describe the relationship between the parameters of the actual shock transition layer and the above-simplified model. Since the ion reflection may be approximated by the specular reflection, the bulk velocities of the upstream and reflected ions in the rest frame of the shock are of the same magnitude but with opposite signs. Therefore, the bulk velocity of the reflected ions as measured in the rest frame of the upstream plasma is twice the shock speed. Since the background electrons and ions correspond to the upstream populations coming into the shock, the ring velocity may be estimated by $V_{r} \sim 2 V_{s}$, where $V_{s}$ is the shock propagation speed defined in the upstream rest frame. If the heating of particles in entering into the shock transition layer and also during the reflection is ignored, the thermal velocity for each component is equal to that in the unperturbed upstream plasma. We thus assume that the background ions and reflected ions have the same thermal spread $\vth{i} = \vth{r}$ for simplicity.

In the following, we denote $\MA = V_{r}/\VA$ and $\MS = V_{r}/\vth{i} = V_{r}/\vth{r}$, which are respectively referred to as the \Alfven and sound Mach numbers. They may correspond, aside from numerical factors of order unity, to the actual Mach numbers of the shock. We expect that non-specular ion reflection and ion heating by the generated turbulence during the reflection process will introduce finite differences between our definitions and the actual shock parameters, but not substantially. Our main purpose is to investigate the dependence on these two major macroscopic shock parameters.

Other important parameters include: the reflected ion density $n_r/n_0$, the electron plasma beta $\beta_{e} = n_e m_e \vth{e}^2/(B^2/8\pi)$, and the electron cyclotron-to-plasma frequency ratio $\Wce/\wpe$, which we call a magnetization. The reflected ion density is typically $n_r/n_0 \sim 0.2{\rm - }0.3$, but may occasionally reach up to $\sim 0.5$ at a high Mach number shock exhibiting periodic self-reformation. The electron plasma beta is of order unity in the solar wind at 1 AU, whereas it varies quite a lot in the interstellar medium. The magnetization is typically very small $\Wce/\wpe \ll 1$ both in the solar wind and interstellar medium. It is, however, very common to use artificially large values $\Wce/\wpe \sim 0.1{\rm -}1$ in PIC simulations to save computational resources. Since the quantity is related to the \Alfven speed via $\VA/c = (\Wce/\wpe) (m_e/m_i)^{1/2}$, typical PIC simulations assume artificially high \Alfven speeds relative to the speed of light. Therefore, high \Alfven Mach number {\it non-relativistic} shocks simulated by a PIC code tend to have substantially high shock speeds relative to the speed of light $V_{s}/c \sim O(0.1)$. Since this is much higher than typical shock speeds in young SNRs, theoretical understanding of the dependence on the magnetization $\Wce/\wpe$ and the ion-to-electron mass ratio $m_i/m_e$ should carefully be taken into account when interpreting the results of PIC simulations adopting unrealistic parameters.

\section{Linear Analysis}
\label{sec:linear}
In the following, we present the result of linear analysis for the periodic model described in the previous section. We restrict ourselves to the wave propagation along the direction of the background magnetic field. The assumption is appropriate for the AIC instability as it is essentially a parallel propagating mode. Probably, the Weibel instability seen in previous 2D PIC simulations \citep{Kato2010,Matsumoto2015,Bohdan2020a} will be correctly reproduced as well. On the other hand, the assumption precludes the mirror instability, which is fundamentally a mode of oblique propagation. As we will see in Section \ref{sec:simulation}, obliquely propagating modes do not seem to be dominant at least in a very high Mach number regime. Therefore, we think that the parallel propagation assumption is an adequate simplification for our purpose to understand the transition of instabilities from low to high Mach numbers. Indeed, we will see below that both the AIC and Weibel instabilities appear as two different limits of the single dispersion relation. Note that this does not mean that the mirror instability is entirely negligible at lower Mach number shocks \citep{Min2017}.

The dispersion relation for circularly polarized electromagnetic waves propagating parallel to the ambient magnetic field is given by the following equation \citep{StixBook}:
\begin{align}
    &D \left( \omega, k \right)
    = 1 - \frac{c^2 k^2}{\omega^2} +
    \\
    &
    \pi \sum_{s} \frac{\omega_{ps}^2}{\omega^2}
    \int_{-\infty}^{+\infty} \int_{0}^{\infty}
    \frac{
        v_{\perp} k \frac{\partial F_{s}}{\partial v_{\para}} +
        \left( \omega - k v_{\para} \right) \frac{\partial F_{s}}{\partial v_{\perp}}
        }{\omega - k v_{\para} - \Omega_{s}}
    v_{\perp}^2 d v_{\perp} d v_{\parallel} = 0,
\label{eq:dispersion}
\end{align}
where $\omega_{ps}$ and $\Omega_{s}$ are respectively the plasma and (signed) cyclotron frequency of particles species $s$. Note that the positive (negative) frequency in the above equation indicates left-hand (right-hand) polarizations, respectively. Our task is to find the solutions $\omega$ as a function of the wavevector $k$, by substituting the model velocity distribution functions, Eq.~(\ref{eq:ring}) for the reflected ions ($s = r$) and Eq.~(\ref{eq:maxwellian}) for the background electrons and ions ($s = e, i$), to the above equation.

It is very well known that Eq.~(\ref{eq:dispersion}) describes the AIC instability driven by a sufficiently high ion temperature anisotropy $T_{\perp}/T_{\para} > 1$ destabilized by the cyclotron resonance: $\omega - k v_{\para} - \Omega_{i} \approx 0$ \citep{Kennel1966}. It is, however, not entirely clear how the Weibel instability in the unmagnetized limit is obtained from the same equation.

\subsection{Cold Plasma Approximation}
\label{sec:cold-plasma}
It is instructive to first look at the basic properties of the dispersion relation in the simplest limit. For this purpose, we consider the cold plasma approximation in the sense that the thermal spread of all the three populations are negligible $\vth{s} = 0$. The dispersion relation in this limit may be obtained by replacing the distribution functions by normalized $\delta$ functions, yielding
\begin{align}
    1 -
    &
    \frac{k^2 c^2}{\omega^2} -
    \frac{\wpe^2}{\omega \left( \omega - \Wce \right)} -
    \frac{\wpi^2}{\omega \left( \omega - \Wci \right)}
    \\ -
    &
    \frac{\wpr^2}{\omega \left( \omega - \Wcr \right)} -
    \frac{k^2 V_{r}^2 }{2 \omega^2}
    \frac{\wpr^2}{\left( \omega - \Wcr \right)^2} = 0.
\label{eq:dispersion-cold}
\end{align}
Notice that the last two terms on the left-hand side are contributions coming from the reflected ions, and obviously $\Wcr = \Wci$.

To consider the limit for the AIC instability, we rewrite Eq.~(\ref{eq:dispersion-cold}) for a low-frequency wave $\omega/k c \ll 1$ and obtain
\begin{align}
    &
    \left[
        \left( \frac{\omega}{\Wci} - 1\right) \left( \frac{k c}{\wpi} \right)^2 +
        \left( \frac{\omega}{\Wci} \right)^2
    \right]
    \left( \frac{\omega}{\Wci} - 1 \right)
    \\
    & = -
    \frac{1}{2} \left( \frac{k c}{\wpi} \right)^2
    \left( \frac{n_r}{n_i} \right)
    \left( \frac{V_{r}}{\VA} \right)^2.
\end{align}
If the right-hand side is negligible, this equation describes the dispersion relation of low-frequency electromagnetic waves in the cold plasma limit (AIC for $\omega > 0$ and whistlers for $\omega < 0$). For a finite right-hand side, we seek an unstable solution close to the ion cyclotron frequency. Substituting $\omega/\Wci = 1 + \delta$ and assuming $\delta \ll 1$, we find an instability \citep{Wu1972a}:
\begin{align}
    \frac{\gamma}{\Wci} = {\rm Im}(\delta) \simeq
    \left( \frac{n_{r}}{2 n_0} \right)^{1/2}
    \MA
    \label{eq:cold-aic}
\end{align}
where we have assumed $n_i \simeq n_0$. Formally, the validity of the above perturbative calculation requires that the \Alfven Mach number should not be too high.

Let us now consider the opposite limit in which the \Alfven Mach number is large $\MA \gg 1$. In other words, the magnetic field is so weak that the cyclotron frequencies of both electrons and ions are negligible compared to the wave frequency $|\omega| \gg |\Wce| \gg \Wci$. The dispersion relation in this case, again for $\omega/kc \ll 1$, simplifies to
\begin{align}
    k^2 c^2 + \wpe^2 + \wpi^2 + \wpr^2 +
    \frac{k^2 V_{r}^2}{2 \omega^2} \wpr^2  = 0,
\label{eq:dispersion-unmag}
\end{align}
which has a purely growing mode approximately given by
\begin{align}
    \frac{\gamma}{\omega_{p0}} = \frac{{\rm Im}(\omega)}{\omega_{p0}} \simeq
    \left( \frac{n_r}{2 n_0} \right)^{1/2}
    \left( \frac{V_{r}}{c} \right)
    \frac{1}{\left( 1 + \wpe^2/k^2 c^2 \right)^{1/2}},
\end{align}
where $\omega_{p0} = (4 \pi n_0 e^2/m_i)^{1/2}$ is the ion plasma frequency defined with the total density $n_0$. It is easy to identify the mode is equivalent to the Weibel instability driven by counter-streaming symmetric beams \citep{Achterberg2007a}. Alternatively, one may formally rewrite the growth rate normalized to the ion cyclotron frequency:
\begin{align}
    \frac{\gamma}{\Wci} \simeq
    \left( \frac{n_r}{2 n_0} \right)^{1/2}
    \MA
    \frac{1}{\left( 1 + \wpe^2/k^2 c^2 \right)^{1/2}},
    \label{eq:cold-weibel}
\end{align}
even though we have taken the limit of weak magnetic field $\Wci \rightarrow 0$, $\MA \rightarrow \infty$.

The factor $(1 + \wpe^2/k^2 c^2)^{1/2}$ appears in the denominator represents an effective screening effect of the ion current by the background electrons \citep{Achterberg2007a}. It is associated with the inductive electric field generated in response to the growing magnetic field via $c \nabla \times {\bm E} = - \partial {\bm B} / \partial t$, which accelerates electrons in such a way that the applied ion current is compensated. It is seen that the screening effect significantly reduces the growth rate at wavelengths longer than the electron inertial length $k c /\wpe \ll 1$, whereas it is negligible at short wavelengths because electrons are no longer able to respond to such fluctuations.

We have seen so far that the AIC and Weibel instabilities are obtained in the two opposite limits: $\gamma/\Wci \ll 1$ for the AIC instability, whereas $\gamma/\Wci \gg m_i/m_e$ (or $\gamma/|\Wce| \gg 1$) for the Weibel instability, respectively. The latter condition is, however, too stringent that will not be satisfied in most of space and astrophysical shocks. We thus investigate the intermediate regime: $1 \ll |\omega|/\Wci \ll m_i/m_e$. This implies that the electrons are essentially magnetized whereas the ions are not. The dispersion relation may then be rewritten as
\begin{align}
    \left( \frac{\omega}{\Wci} \right)^3 +
    \left[ 1 + \left( \frac{k \VA}{\Wci} \right)^2 \right]
    \left( \frac{\omega}{\Wci} \right)^2 +
    \left( \frac{k V_{r}}{\Wci} \right)^2 \left( \frac{n_r}{2 n_0} \right) = 0.
\end{align}
It is useful to consider approximate unstable solutions which may easily be obtained for short and long wavelength limits:
\begin{subnumcases} {\label{eq:cold-intermediate} \frac{\omega}{\Wci} \simeq}
    i \left( \frac{n_r}{2 n_0} \right)^{1/2} \MA
    & \hspace{-0.8cm} 
    $\displaystyle \quad \left( \frac{k \VA}{\Wci} \right)^2 \gg \frac{|\omega|}{\Wci}$,
    \nonumber \\
    \label{eq:cold-intermediate-highk} \\
    \frac{1 + \sqrt{3} i}{2}
    \left( \frac{n_r}{2 n_0} \right)^{1/3}
    \left( \frac{k \VA}{\Wci} \right)^{2/3} \MA^{2/3}
    & \hspace{-0.8cm} 
    $\displaystyle \quad \left( \frac{k \VA}{\Wci} \right)^2 \ll \frac{|\omega|}{\Wci}$.
    \nonumber \\
    \label{eq:cold-intermediate-lowk}
\end{subnumcases}

It is interesting to note that the growth rate in the short wavelength limit Eq.~(\ref{eq:cold-intermediate-highk}) is formally identical to the AIC instability Eq.~(\ref{eq:cold-aic}), and also to the Weibel instability Eq.~(\ref{eq:cold-weibel}) in the absence of the electron screening effect (or $k c/\wpe \gg 1$). The short wavelength limit has a negligible real frequency (in the sense ${\rm Im}(\omega) \gg {\rm Re}(\omega)$) and is recognized as a Weibel-like instability. The screening effect by the background electrons disappears, and the growth rate is larger than in an unmagnetized plasma because of the magnetization of electrons. In other words, the inductive electric field now causes an $E \times B$ drift for the magnetized electrons perpendicular to the electric field, which prohibits them from screening the applied ion current.

On the other hand, the long-wavelength limit has both real and imaginary parts of the same order of magnitude. The positive real part (or left-hand polarization) for the unstable mode shares the same property with the AIC instability. Numerical solutions for the dispersion relation Eq.~(\ref{eq:dispersion-cold}) confirm that the unstable modes in the short and long wavelength regimes are well represented by Eq.~(\ref{eq:cold-intermediate}), which are then smoothly connected with each other.

In the following, we will distinguish the instability by the magnitude of $\gamma/\Wci$. The classical AIC instability may be identified by the condition $\gamma/\Wci \lesssim 1$, whereas the instability with $\gamma/\Wci \gg 1$ will be referred to as the Weibel instability. We have confirmed through an extensive parameter survey that this distinction correctly characterizes the instability property.

\subsection{Solutions for Hot Plasma}
\label{sec:hot-plasma}
Having seen analytic expressions for some limiting cases with the cold plasma approximation, we now look into numerical solutions for the fully kinetic dispersion relation Eq.~(\ref{eq:dispersion}) to discuss relations between these limits as well as modifications introduced by the kinetic effect. It is well known that the response of Maxwellian components in Eq.~(\ref{eq:dispersion}) can be expressed analytically using the plasma dispersion function \citep{StixBook}. Although some analytic formulae are available for the integration in perpendicular velocity for the ring distribution Eq.~(\ref{eq:ring}) \citep{Vandas2015}, we evaluated it by numerical integration. Following the approach adopted by \citet{Umeda2012c}, we use the Simpson formula for the evaluation of the integral. Note that the integration of the ring distribution in parallel velocity can be written by the plasma dispersion function as well. Once the response of each component is obtained, root finding can be performed in the same way as in standard kinetic plasma dispersion analysis.

\subsubsection{Dependence on \Alfven Mach Number}
\label{sec:hot-Ma-dep}

\begin{figure}[tb]
    \centering
    \includegraphics[height=0.7\textheight,width=0.5\textwidth,keepaspectratio]{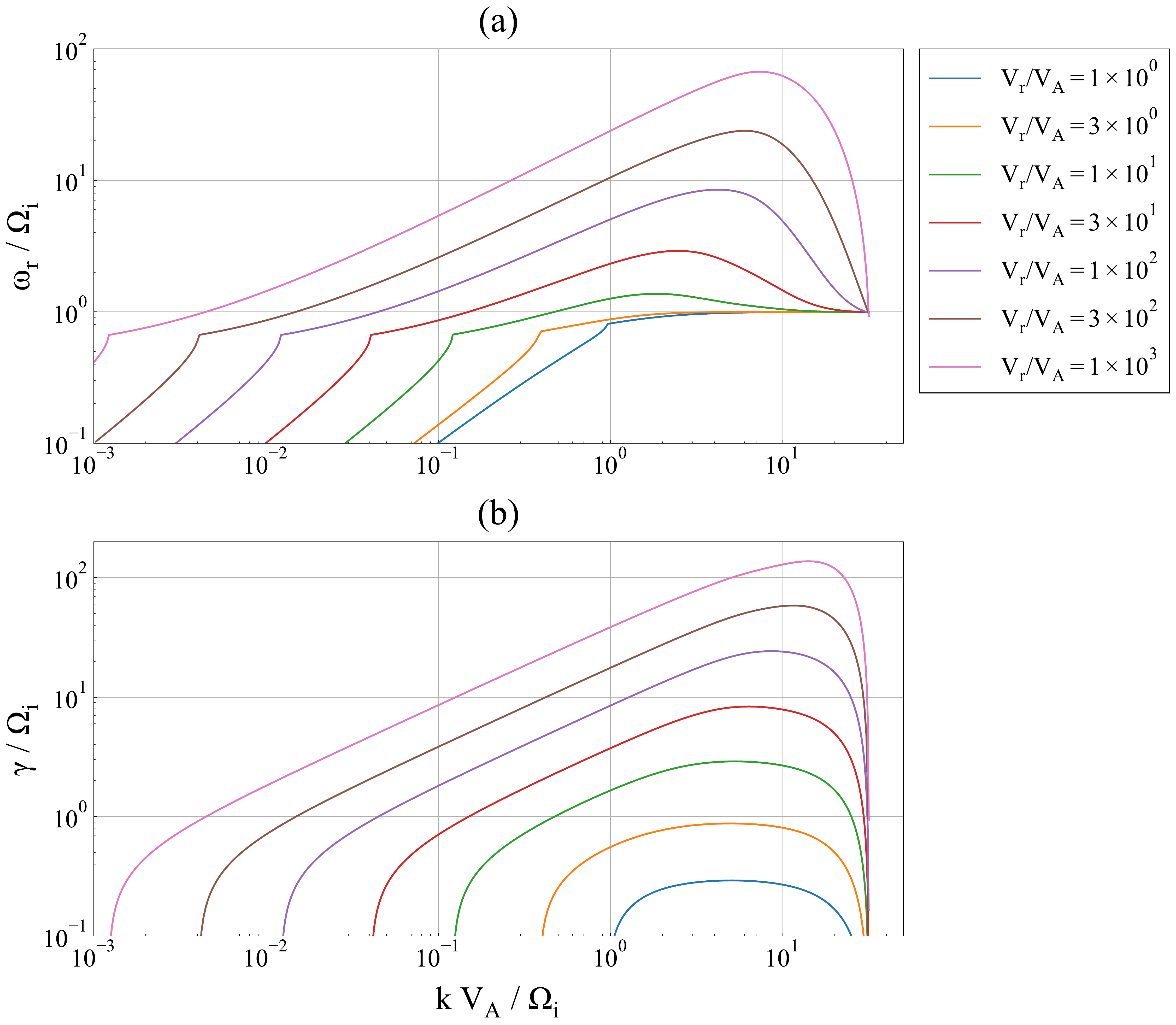}
    \caption{
        \Alfven Mach number dependence of unstable solution. (a) real, and (b) imaginary parts of the numerical solutions for $\MA = V_r/\VA = 1 \times 10^{0}, 3 \times 10^{0}, 1 \times 10^{1}, 3 \times 10^{1}, 1 \times 10^{2}, 3 \times 10^{2}, 3 \times 10^{2}, 1 \times 10^{3}$ are shown. The other parameters are:  $n_r/n_0=0.2$, $\MS = 100$, $m_i/m_e = 1836$, $\Wce / \wpe = 5 \times 10^{-3}$, $\beta_e =10$.
    }
    \label{fig:Ma-dep}
\end{figure}

First, we investigate the dependence on the \Alfven Mach number. Fig.~\ref{fig:Ma-dep} shows the real and imaginary parts of the numerical solution as a function of wavenumber in the range $1 \leq \MA \leq 10^{3}$, with all the other parameters fixed: $n_r/n_0=0.2$, $\MS = 100$, $m_i/m_e = 1836$, $\Wce / \wpe = 5 \times 10^{-3}$, $\beta_e =10$. At small $\MA$, the peak growth rate is lower than the ion gyrofrequency, whereas the real frequency is nearly equal to it. This is consistent with the analytical formula Eq.~(\ref{eq:cold-aic}) for the AIC instability. The growth rate increases as the \Alfven Mach number increases and becomes larger than the ion gyrofrequency at $\MA \gtrsim 10$. The real part for this regime also increases beyond the ion gyrofrequency and is of the same order as the growth rate. We have confirmed that the long-wavelength part of the solutions for $\MA \gg 10$ is well represented by Eq.~(\ref{eq:cold-intermediate-lowk}). The growth rate even in this regime is still much lower than the electron gyrofrequency $\gamma/\Wci \ll m_i/m_e$ and is consistent with the assumption made for obtaining the intermediate regime Eq.~(\ref{eq:cold-intermediate}). Therefore, we understand that the AIC instability at lower \Alfven Mach numbers ($\MA \lesssim 10$) continuously transitions to the Weibel instability at very high Mach numbers ($\MA \gg 10$).

Note that the cold plasma theory predicts that the growth rate asymptotes to Eq.~(\ref{eq:cold-intermediate-highk}) in the short wavelength limit. On the other hand, the kinetic growth rate decreases rapidly at a wavelength of $k \VA/\Wci \sim 30$ regardless of $\MA$. This may be understood as a kinetic damping effect introduced by the finite thermal spread of the reflected ion component. To see this, let us first consider an instability with sufficiently high $\MA$. The instability is driven by the reflected ions, which primarily travel in the plane perpendicular to the background magnetic field. Nevertheless, they also travel along the magnetic field with a small thermal velocity $\vth{r}$ for a distance of $\vth{r}/\gamma$ during the instability growth time  $1/\gamma$. The instability requires the driving particle population must be coherent in the sense that $\vth{r}/\gamma \ll 1/k$, because otherwise, the particles experience many different phases of the perturbation during the wave growth time. The instability will thus be suppressed beyond a critical wavenumber $k_{*} \sim \gamma/\vth{r}$. Substituting the analytic expression for the growth rate Eq.~(\ref{eq:cold-intermediate-highk}) in the short wavelength regime, we obtain
\begin{align}
    \frac{k_{*} \VA}{\Wci} \sim
    \left( \frac{n_r}{2 n_0} \right)^{1/2}
    \MS.
    \label{eq:critical-k}
\end{align}
This estimate is independent on $\MA$ and gives $k_{*} \VA/\Wci \sim 30$ for $n_r/n_0 = 0.2$ and $\MS = 100$, which is roughly consistent with the numerical results. Notice that the condition for the kinetic damping may be understood in terms of cyclotron resonance condition $\omega = k v_{\para} - \Wci$ with $\omega \sim \gamma \gg \Wci$ and $v_{\para} \sim \vth{r}$.

\subsubsection{Dependence on Sound Mach Number}
\label{sec:hot-Ms-dep}

\begin{figure}[tb]
    \centering
    \includegraphics[height=0.7\textheight,width=0.5\textwidth,keepaspectratio]{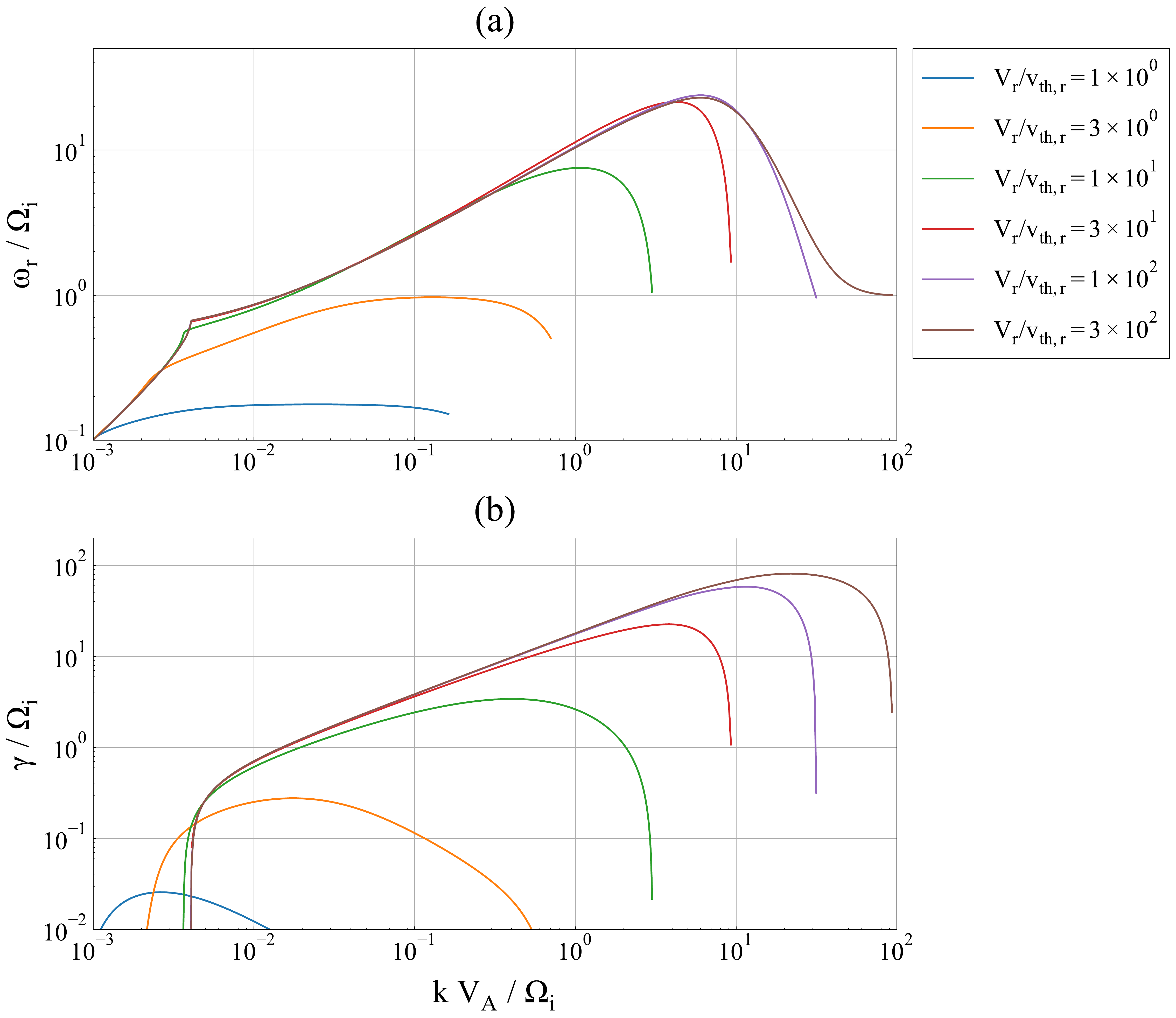}
    \caption{
        Sound Mach number dependence of unstable solution. (a) real, and (b) imaginary parts of the numerical solutions for $\MS = V_r/\vth{r} = 1 \times 10^{0}, 3 \times 10^{0}, 1 \times 10^{1}, 3 \times 10^{1}, 1 \times 10^{2}, 3 \times 10^{2}, 3 \times 10^{2}$ are shown. The other parameters are: $n_r/n_0 = 0.2$, $\MA = 300$, $m_i/m_e = 1836$, $\Wce / \wpe = 5 \times 10^{-3}$, $\beta_e =10$.
    }
    \label{fig:Ms-dep}
\end{figure}

As we have seen above, a finite thermal velocity of the reflected ions may play a significant role, in particular at short wavelength $k \gtrsim k_{*}$ even for the strongly driven Weibel regime $\MA \gg 10$. To investigate the dependence in more detail, we calculate the dispersion relation for a broad range of sound Mach numbers with keeping the other parameters constant. Since we normalize the velocity by the \Alfven speed and $\MA$ is fixed, we change only the thermal velocity of ions (with always assuming $\vth{r} = \vth{i}$). The results are shown in Fig.~\ref{fig:Ms-dep} for $n_r/n_0 = 0.2$, $\MA = 300$, $m_i/m_e = 1836$, $\Wce / \wpe = 5 \times 10^{-3}$, $\beta_e =10$. As predicted by Eq.~(\ref{eq:critical-k}), the increase (decrease) of the sound Mach number tends to shift the maximum growth wavenumber toward higher (lower) wavenumbers for $\MS \gg 1$. Correspondingly, the maximum growth rate becomes larger for higher $\MS$. On the other hand, the maximum growth rate and the real frequency drop down below the gyrofrequency when $\MS \sim 3$. In this case, the instability characteristics are essentially the same as the AIC instability.

The transition from the Weibel instability at $\MS \gg 1$ to the AIC instability at $\MS \sim 3$ may qualitatively be understood as follows. The critical wavenumber $k_{*}$ gradually decreases as decreasing $\MS$, and at some point $k_{*} \VA/\Wci \sim 1$ will be reached. This condition formally violates the assumptions of $\gamma/\Wci \gg 1$ and $(k \VA/\Wci)^2 \gg |\omega|/\Wci$, both of which must be satisfied to obtain the growth rate Eq.~(\ref{eq:cold-intermediate-highk}). This suggests that the instability at $k_{*} \VA/\Wci \sim 1$, or equivalently $\MS \sim (n_r/2 n_0)^{-1/2} \sim 3$ from Eq.~(\ref{eq:critical-k}), is no longer considered as the Weibel instability. The numerical solutions confirm this estimate is roughly correct.

Qualitatively speaking, the instability becomes Weibel-like only in cases where both \Alfven and sound Mach numbers are sufficiently large. The instability becomes AIC-like if either of them does not satisfy some threshold.

\subsubsection{Summary}
\label{sec:hot-summary}

We have also investigate the dependence on $n_r/n_0$, $\beta_e$, $m_i/m_e$, $\wpe/\Wce$ (see, Appendix \ref{sec:appendix}) and found that they do not change the qualitative property of instability. Conventionally, the Weibel instability has been analyzed for an unmagnetized plasma. For applications to collisionless shocks, the unmagnetized plasma assumption implies the \Alfven Mach number is formally infinite, and the growth rate should be written only in terms of $V_s/c$, i.e., the shock speed relative to the speed of light. This makes it difficult to understand which of the two parameters, either $V_s/\VA$ or $V_s/c$, is controlling the instability. The fact that the instability property driven by the magnetized ring distribution is simply determined by the \Alfven and sound Mach numbers clearly indicates that $V_s/c$ is not a relevant parameter, as is clear from the relationship $V_s/c = \MA (\wpe/\Wce)^{-1} (m_i/m_e)^{1/2}$. Therefore, the Weibel instability can be a dominant mode even at non-relativistic shocks, such as young SNR shocks with typical shock speeds of order $V_s/c \sim O(0.01)$, if the \Alfven and sound Mach numbers are sufficiently high.

\begin{figure}[H]
    \centering
    \includegraphics[height=0.7\textheight,width=0.5\textwidth,keepaspectratio]{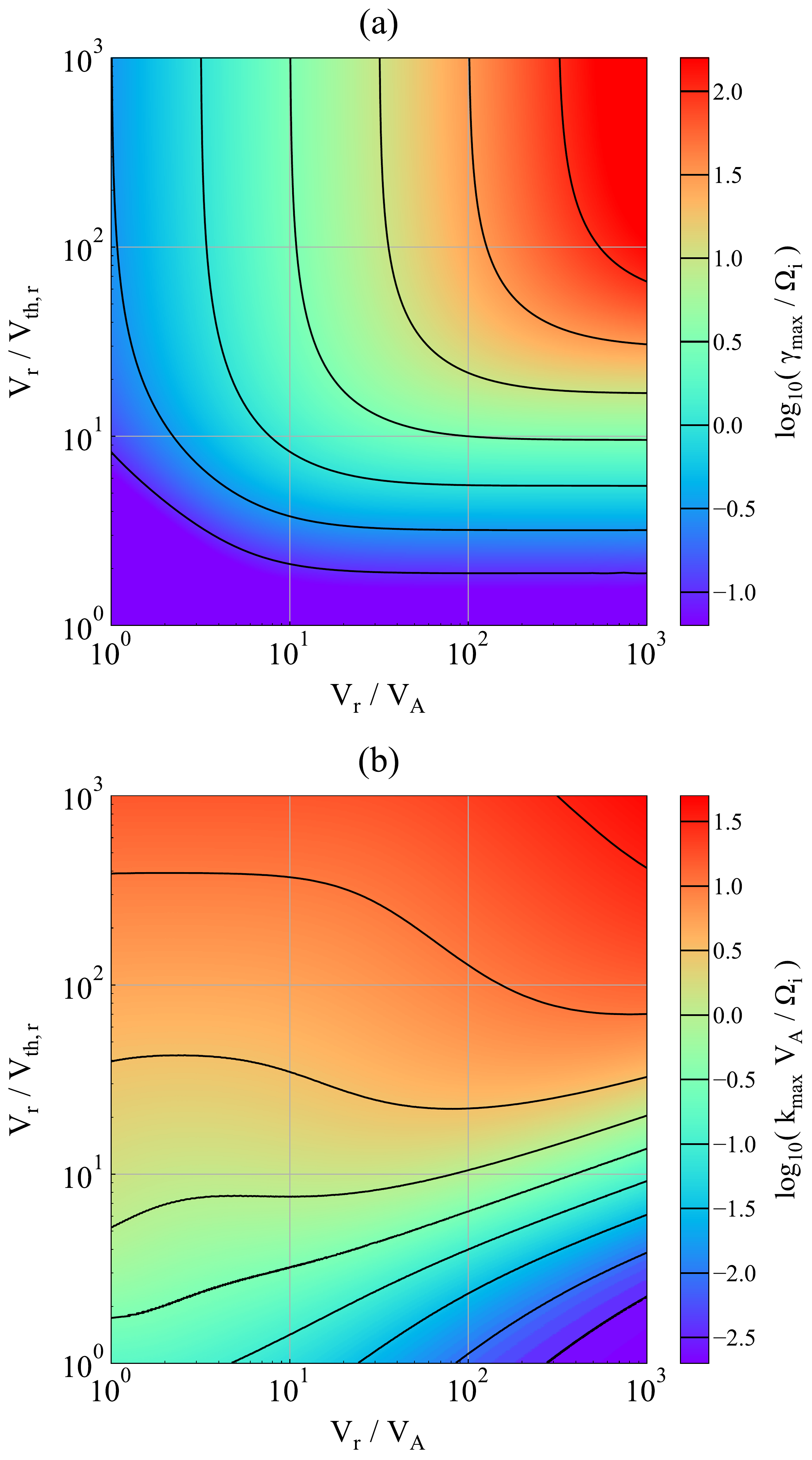}
    \caption{
        Summary plots for the unstable solution in the fully kinetic model: (a) the maximum growth rate, (b) the wavenumber at the maximum growth. The parameters are $n_r/n_0 = 0.2$, $m_i/m_e = 1836$, $\Omega_e / \omega_{pe} = 5 \times 10^{-3}$, and $\beta_e = 0.1$.
    }
    \label{fig:hot-summary}
\end{figure}

As we now understand that the instability property is regulated mostly by the two Mach numbers $\MA$ and $\MS$ with relatively minor modifications introduced by other parameters, a summary plot representing the instability property in a plane spanned by the two parameters will be useful. Fig.~\ref{fig:hot-summary} provides such plots for the maximum growth rate $\gamma_{\rm max}$ on the unstable branch, and the wavenumber at the maximum growth $k_{\rm max}$ represented in contours. The other parameters used are $n_r/n_0 = 0.2$, $m_i/m_e = 1836$, $\Wce/\wpe = 5 \times 10^{-3}$, $\beta_{e} = 0.1$.

It is clear that the growth rate increases when both of the two parameters are increased. We see that $k_{\rm max}$ shows a stronger dependence on the sound Mach number than the \Alfven Mach number, which may be understood as the effect of kinetic damping. We may expect that a strong Weibel-dominated shock will be realized for $\gamma_{\rm max}/\Wci \gtrsim 10$ so that the instability can grow in a time scale much shorter than the dynamical time scale of the shock. The linear theory thus predicts that \Alfven and sound Mach numbers greater than $\sim 20-40$ will be required for a shock to behave as Weibel-dominated; otherwise less dramatic shock-surface rippling will be the dominant mode.

\section{Particle-In-Cell Simulation}
\label{sec:simulation}

\subsubsection{Simulation Setup}
\label{sec:setup}
We carried out 2D PIC simulations with the same homogeneous model consisting of the three particle populations (background electrons, ions, and reflected ions) that is used in the linear dispersion analysis. The simulations were performed in the $x{\rm -}y$ plane with the background magnetic field parallel to $y$ direction.

Our main interest in performing PIC simulations is for $\gamma_{\rm max}/\Wci \gg 1$, or the Weibel-like instability regime. The reason for this is that the model is, strictly speaking, not appropriate for any phenomena with the time scale longer than $\Wci^{-1}$ because the upstream plasma flowing with the speed $V_{s}$ over the typical shock thickness $V_{s}/\Wci$ has a finite interaction time with the shock of order $\Wci^{-1}$. The AIC and mirror instabilities with $\gamma_{\rm max}/\Wci \lesssim 1$ must be strongly affected by the inhomogeneity of the shock, which is indeed the reason why it has been difficult to identify the relevant instability for the rippling \citep{Winske1988}. In addition, there have been substantial works that investigate AIC and mirror instabilities in a homogeneous plasma \citep{Gary1997,Shoji2009,Yoon2017}. On the other hand, the Weibel instability in a magnetized plasma has yet remained relatively unexplored.

The simulation box was a square with a side length of approximately $5 \, \VA/\Wci$ (with an exception that will be discussed later in Section \ref{sec:reconnection}). The box size was sufficient to contain at least a few of the fastest-growing modes for the parameter regime of interest. The time integration was typically performed up to $\sim \Wci^{-1}$, which we found sufficient to capture the saturation of the fastest growing modes. We used a fixed grid size equal to the electron Debye length $\lambda_{{\rm D},e} = \vth{e} / \wpe$ for all the simulations presented in this paper. The time step is limited by the CFL (Courant-Friedrichs-Lewy) condition with respect to the speed of light. We determined the time step by the condition $c \Delta t / \Delta x = 1$ as long as $\wpe \Delta t < 0.1$, otherwise a time step of $\wpe \Delta t = 0.1$ was used. The number of particles per cell was 50 at the initial condition for all the species.

A simulation run may be characterized by the following parameters: $n_r/n_0$, $\MA$, $\MS$, $\Wce/\wpe$, $m_i/m_e$, $\beta_e$. Note that we always assumed $\vth{i} = \vth{r}$ for the simulations as well. Parameters used in the survey presented in this paper are summarized in Table \ref{tab:parameters}.

\begin{table}[htbp]
    \centering
    \caption{Simulation Parameters.}
    \label{tab:parameters}
    \begin{ruledtabular}
        \begin{tabular}{ccccccc}
            Run & $n_r/n_0$ & $\MA$ & $\MS$ & $\Wce/\wpe$ & $m_i/m_e$ & $\beta_e$ \\
            \hline
            1   & 0.2       & 50  & 100 & 0.05        & 100     & 8       \\
            2   & 0.2       & 50  & 100 & 0.05        & 100     & 2       \\
            3   & 0.05      & 50  & 100 & 0.05        & 100     & 8       \\
            4   & 0.4       & 50  & 100 & 0.05        & 100     & 8       \\
            5   & 0.2       & 25  & 100 & 0.05        & 100     & 8       \\
            6   & 0.2       & 100 & 100 & 0.05        & 100     & 8       \\
            7   & 0.2       & 50  & 25  & 0.05        & 100     & 8       \\
            8   & 0.2       & 50  & 400 & 0.05        & 100     & 8       \\
            9   & 0.2       & 50  & 100 & 0.025       & 100     & 8       \\
            10  & 0.2       & 50  & 100 & 0.1         & 100     & 8       \\
            11  & 0.2       & 50  & 100 & 0.05        & 25      & 8       \\
            12  & 0.2       & 50  & 100 & 0.05        & 400     & 8       \\
            13  & 0.5       & 100 & 500 & 0.05        & 100     & 8       \\
        \end{tabular}
    \end{ruledtabular}
\end{table}

\subsubsection{Role of Electron-Scale Instabilities}
\label{sec:eletron-insta}
The ring distribution of the reflected ion component can also drive rapidly growing instabilities at the electron scale via the interaction with the background electrons \citep{Wu1984a}. These include Buneman instability, electron cyclotron drift (ECD) instability, and modified two-stream (MTS) instability. To understand this, we should first note that the linear stability of the system against these modes is determined by a reduced one-dimensional distribution function. For instance, in our coordinate system, electrostatic waves with $k$ vector parallel to $x$ direction is determined by $\int \int f(\bm{v}) \, d v_{y} d v_{z}$, i.e., the distribution function integrated over $v_{y}$ and $v_{z}$. A relatively cold ring distribution will appear as symmetric counter-streaming beams in the reduced distribution in $v_{x}$. Therefore, the ring distribution of the reflected ions drives various kinds of electron-scale instabilities with the growth rates much higher than that of the ion-scale mode. The question is whether or not such rapidly growing modes affect the much slower instability at the ion scale.

\begin{figure}[tb]
    \centering
    \includegraphics[height=0.7\textheight,width=0.5\textwidth,keepaspectratio]{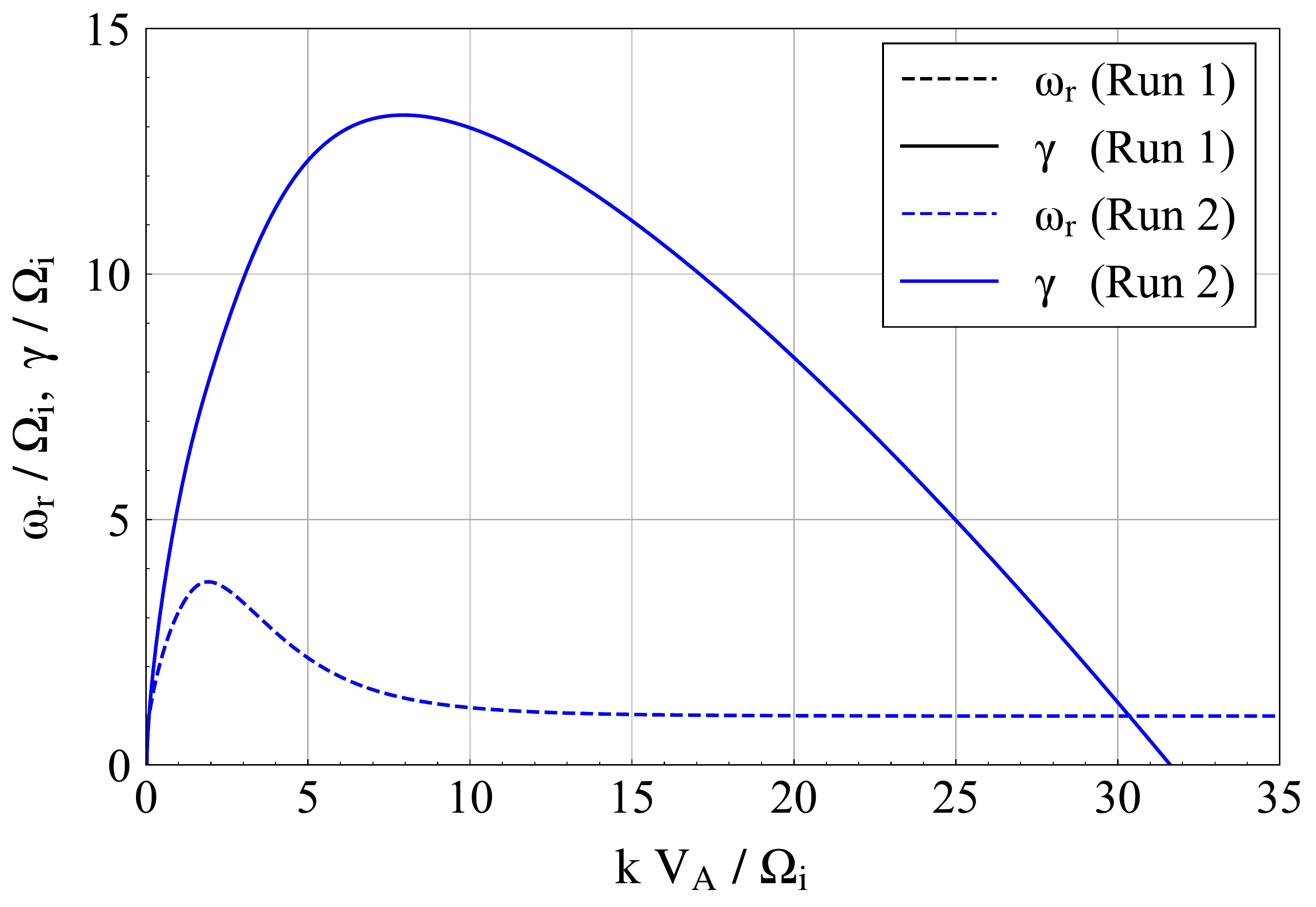}
    \caption{
        Linear dispersion relations for Run 1 ($\beta_e = 8$) and Run 2 ($\beta_e = 2$). The solid and dashed lines respectively show the imaginary and real parts of the solutions. The solutions for Run 1 and 2 are shown in black and blue, respectively. The two solutions are indistinguishable from each other.
    }
    \label{fig:linear-simulation}
\end{figure}

\begin{figure}[htb]
    \centering
    \includegraphics[height=0.7\textheight,width=0.5\textwidth,keepaspectratio]{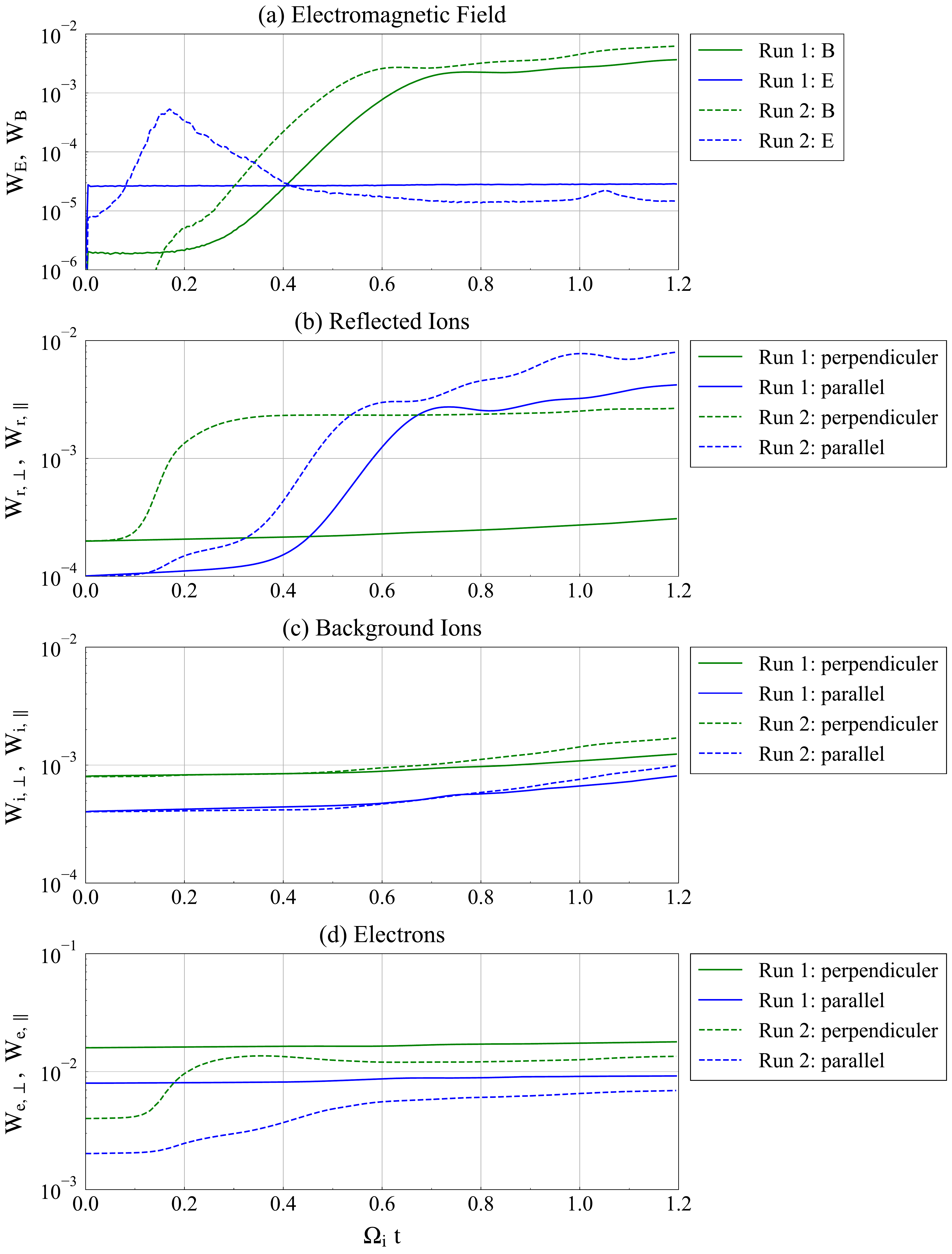}
    \caption{
        Energy history for Run 1 and 2: (a) electric and magnetic field energies, (b) reflected ion thermal energies, (c) background ion thermal energies, (d) electron thermal energies. The solid and dashed lines respectively show the results of Run 1 ($\beta_e = 8$) and Run 2 ($\beta_e = 2$). The energy is normalized to the initial kinetic energy of the reflected ions $n_r m_i V_r^2 / 2$.
    }
    \label{fig:summary-run12}
\end{figure}

To answer the question, we compare simulations with two different initial electron temperatures with all the other parameters being identical: Run 1 with $\beta_e = 8$ and Run 2 with $\beta_e = 2$. Fig.~{\ref{fig:linear-simulation}} shows the theoretical dispersion relations of the electromagnetic mode corresponding to the two cases, which are nevertheless indistinguishable from each other. We thus understand any differences between the two runs arise from the differences during an early phase where the electron-scale instabilities are dominant. Fig.~\ref{fig:summary-run12} shows the time evolution of the electric and magnetic field energies $W_E = \bm{E}^2/8\pi$, $W_B = (\bm{B}^2 - \bm{B}^2_0)/8\pi$ and parallel and perpendicular thermal energies $W_{s,\parallel}, W_{s,\perp}$ of the three particle populations for the two runs, which are all normalized to the initial reflected ions kinetic energy $n_r m_i V_r^2 / 2$. The thermal energies are defined by the second order velocity moment. Note that the perpendicular thermal energy of the reflected ions is defined by
\begin{align}
    W_{r,\perp} &
    =
    \int \int v_{\perp}^2 f(v_{\perp}, v_{\parallel}) \, 2 \pi v_{\perp} d v_{\perp} d v_{\parallel} 
    \\ &
    -
    \frac{1}{n_r} \left( 
        \int \int v_{\perp}   f(v_{\perp}, v_{\parallel}) \, 2 \pi v_{\perp} d v_{\perp} d v_{\parallel}
    \right)^2.
\end{align}
This definition approximately gives the energy associated with the perpendicular thermal spread with respect to the ring velocity $V_r$ for the ring distribution Eq.~(\ref{eq:ring}), if the thermal spread is sufficiently small $V_r \gg \vth{r}$.

It is clear that the increased $\beta_e$ (or the electron temperature) in Run 1 suppressed the growth of electrostatic wave activity seen in Run 2 in the early phase $\Wci \, t \lesssim 0.2$. We have confirmed that the electrostatic waves in Run 2 were generated by the Buneman instability, which was stabilized in Run 1 by an enhanced electron Landau damping with the higher electron temperature. Note that, at a relatively high magnetization (i.e., larger $\Wce/\wpe$), a finite background magnetic field becomes important, and the instability occurs now on the electron Bernstein waves, which is called the ECD instability. Nevertheless, the general trend that the instability suffers stabilization by the electron Landau damping remains unchanged even in this case. As a result of the electrostatic wave activity, both the reflected ions and electrons were heated primarily in the direction perpendicular to the ambient magnetic field. We found (not shown) that the ring velocity (i.e., perpendicular bulk velocity) of the reflected ion component has remained almost unaffected by the electron-scale instabilities. This result is consistent with earlier simulations with symmetric counter-streaming beams \citep{Ruyer2015}.

The growth of magnetic field amplitudes in the later stage appeared very similar between the two runs (albeit at different timings), indicating that the impact of the electron-scale instability was not significant. As we have seen, rapidly growing instabilities will increase the temperatures of the particle populations actively involved in them before the ion-scale instability sets in. We may thus understand that the slowly evolving instability may grow from an initial condition modified by the heating in the early phase. Roughly speaking, the electron heating associated with beam instabilities results in the marginal stability condition. Since the electron temperature has no significant effect on the linear growth rate of the ion-scale mode, the effect of electron heating will be minimal. The effect of ion heating may be understood in a similar fashion. Recall that the reflected ion heating occurs primarily in the perpendicular direction. On the other hand, the critical wavenumber Eq.~(\ref{eq:critical-k}) should be defined physically with the parallel thermal velocity, which does not change much in the early phase. Therefore, the growth of the ion-scale instability is not sensitive to the presence/absence of heating associated with the electron-scale instabilities. Note that the reflected ion heating in the parallel direction was significant in the late stage and was clearly associated with the ion-scale instability.

We should also mention the possibility of the MTS instability, which may also be driven by the reflected ions to generate obliquely propagating low-frequency whistler-mode waves. We have not seen any evidence for the instability signature in our simulation runs. It is consistent with the linear theory \citep{Matsukiyo2003b} that predicts $\MA \lesssim  (m_i/m_e)^{1/2} \, \cos \theta_{kB} \, /2$ (where $\theta_{kB}$ is the wave propagation angle with respect to the background magnetic field) is needed for the mode to be destabilized. In addition, it also suffers a significant electron Landau damping at finite $\beta_e$ unless $\cos \theta_{kB}$ is small. Therefore, the MTS instability will not play a role at high Mach number shocks $\MA \gtrsim (m_i/m_e)^{1/2}$ where the ion-scale instability will likely behave as Weibel-like.

\subsubsection{Development of Ion-scale Instability}
\label{sec:ion-insta}
\begin{figure}[tb]
    \centering
    \includegraphics[height=0.9\textheight,width=0.5\textwidth,keepaspectratio]{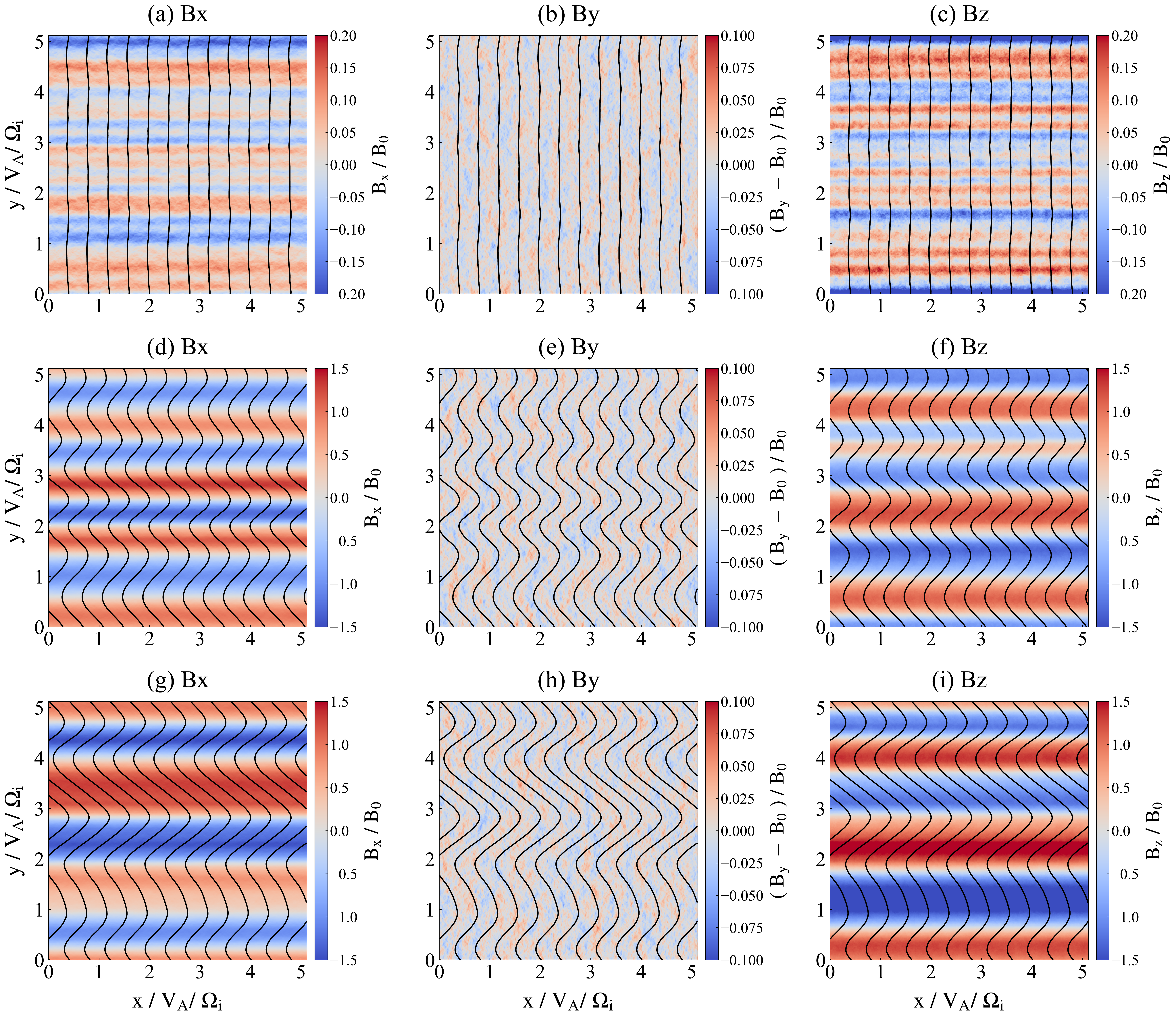}
    \caption{
        Snapshots of the magnetic field for Run 1. Top (a-c) , middle (d-f), bottom (g-i) panels indicate snapshots taken at $\Wci \, t = 0.4, 0.8, 1.2$, respectively. The solid black lines indicate the magnetic field lines.
    }
    \label{fig:snapshots-bfield}
\end{figure}

Fig.~\ref{fig:snapshots-bfield} shows three snapshots of the magnetic field at $\Wci \, t = 0.4, 0.8, 1.2$ for Run 1. We see clearly that the waves were predominantly propagating parallel to the background magnetic field. There were only small-amplitude compressional perturbations that should appear in $\delta B_y$. These observations suggest that the obliquely-propagating mirror instability is not the major mode. We have confirmed essentially the same result for all the simulations in Table \ref{tab:parameters}, whereas we have found signatures of compressional oblique modes at relatively lower Mach numbers (not shown).

\begin{figure}[tb]
    \centering
    \includegraphics[height=0.7\textheight,width=0.5\textwidth,keepaspectratio]{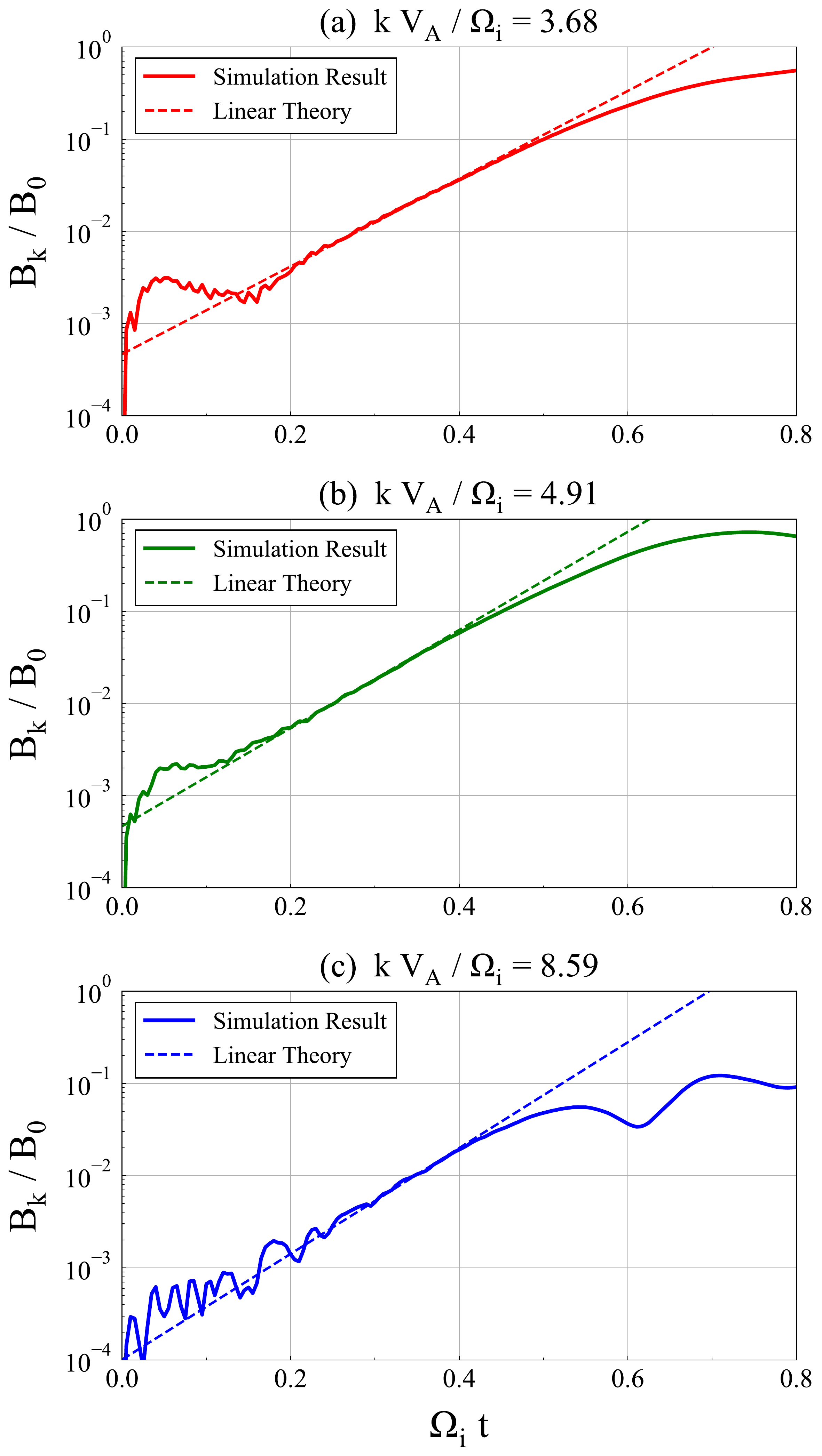}
    \caption{
        Growth of three Fourier amplitudes obtained with Run 1: (a) $k \VA/\Wci = 3.68$, (b) $k \VA/\Wci = 4.91$, (c) $k \VA/\Wci = 8.59$. The solid and dashed lines in each panel indicate the simulation result and the prediction by linear theory.
    }
    \label{fig:linear-growth}
\end{figure}

\begin{figure}[tb]
    \centering
    \includegraphics[height=0.7\textheight,width=0.5\textwidth,keepaspectratio]{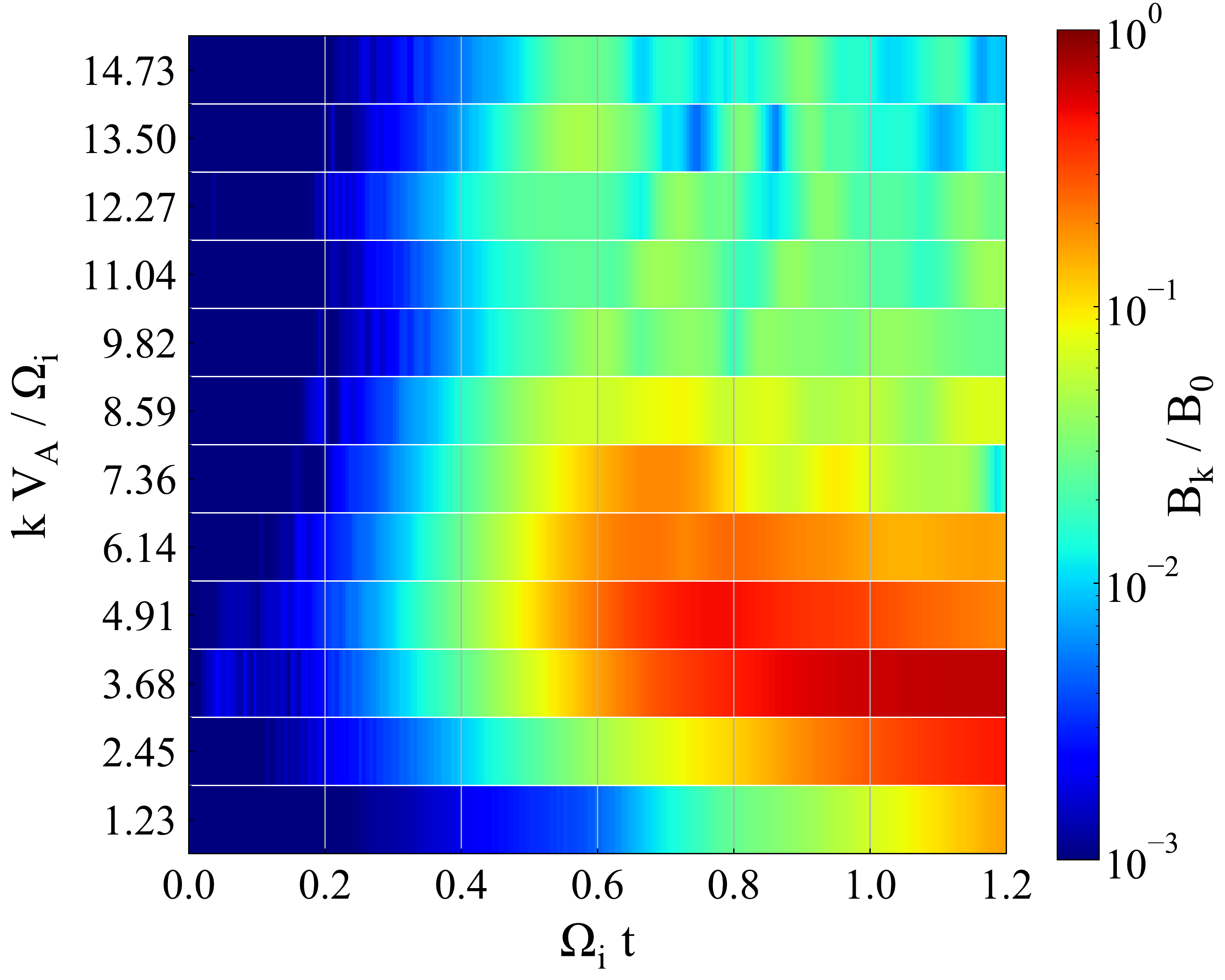}
    \caption{
        Time evolution of wavenumber spectrum for Run 1. Color represents $B_k/B_0$ as a function of parallel wavenumber $k$.
    }
    \label{fig:ky-spectrum}
\end{figure}

We now present a detailed analysis of the dominant parallel propagating modes. Fig.~\ref{fig:linear-growth} shows the growth of fluctuating magnetic field amplitude for three Fourier modes as a function of time. To calculate the mode amplitudes, we average the magnetic field in $x$ direction. With the averaged magnetic field, we construct a complex quantity $\tilde{B} = B_x + i B_z$ and take Fourier transform in $y$ direction to obtain $B_k$, the time development of a particular mode amplitude, for proper comparison with the linear theory. It is clearly seen that the growth curves (solid lines) are consistent with the predictions by the linear theory (dashed lines) for each panel.

The growth of the instability will eventually cease when the mode grows into sufficiently large amplitude so that a nonlinear stabilization effect comes in. Fig.~\ref{fig:ky-spectrum} shows the time development of the parallel wavenumber spectrum for $B_k/B_0$, in which one can observe the saturation level dependence on wavenumber. Note that we take an average of positive and negative wavenumbers in calculating the spectrum utilizing the symmetry in $k$. Apparently, short-wavelength modes saturated earlier at lower levels of fluctuations. The largest amplitude in the nonlinear phase was achieved at $k \VA/\Wci = 3.68$, which did not correspond to the maximum growth wavenumber $k_{\rm max} \VA/\Wci \simeq 8$ (see, Fig.~\ref{fig:linear-simulation}). This may be qualitatively understood in terms of quasi-linear theory with the help of the critical wavenumber $k_{*}$ defined by Eq.~(\ref{eq:critical-k}). As the instability develops, perturbations of the reflected ion velocities by the wave electromagnetic field increase in time. The increased oscillation amplitude of particle velocities may be seen as an increase in apparent temperature, i.e., the effective thermal spread integrated over the simulation box is increased. The quasi-linear heating of particles affects the growth of short-wavelength modes because the critical wavenumber becomes smaller. Therefore, short-wavelength modes will saturate relatively earlier even though the growth rates calculated with the initial condition are higher.

As we will see below, on the other hand, the final saturation of the instability at lower wavenumbers may be controlled by a different mechanism because the wave will evolve into fully nonlinear amplitudes $\delta B/B_0 \sim 1$ before the quasi-linear stabilization becomes significant.

\subsubsection{Nonlinear Saturation Mechanism}
\label{sec:saturation}
Saturation mechanisms of the Weibel instability have been discussed extensively in the past. One of the earliest studies by \citet{Davidson1972} suggested that the saturation may be explained by the particle trapping by the wave magnetic field. The particle trapping mechanism is actually one of the classical models that has been successful in explaining the saturation of kinetic plasma instabilities. When the wave grows into a sufficiently large amplitude, some particles may be trapped within an effective wave potential and perform periodic oscillations around the potential minimum. It is natural to expect that the wave growth will be strongly affected when the wave growth time $1/\gamma$ and the particle bounce period in the potential become comparable with each other because the perturbation onto the particle motion is no longer negligible. In other words, one may predict that the saturation will occur when the following condition is met:
\begin{align}
    \gamma (k) \sim \omega_{\rm b} =
    \left( \frac{e}{m_i} k \frac{V_r}{c} B_k \right)^{1/2},
    \label{eq:trapping-cond}
\end{align}
where $\omega_{\rm b}$ is the magnetic bounce frequency \citep{Davidson1972}. Since the bounce frequency is proportional to square root of the wave amplitude $B_k$, it becomes increasingly more important as the wave grows in amplitude.

We now introduce the energy conversion rate $\varepsilon_B$ as a measure of the saturation level:
\begin{align}
    \varepsilon_{B} (k) = \frac{B_k^2/8 \pi}{n_0 m_i V_r^2/2}.
\end{align}
The trapping condition Eq.~(\ref{eq:trapping-cond}) may be rewritten as follows
\begin{align}
    \varepsilon_{B} (k) \sim
    \left( \frac{\gamma(k)}{\Wci} \right)^{4}
    \MA^{-4}
    \left( \frac{k \VA}{\Wci} \right)^{-2},
    \label{eq:epsilon-trapping}
\end{align}
which gives a theoretical prediction for the saturation as a function of wavenumber. We will call this model as the trapping limit. Note that this estimate may apply to different wavenumbers independently. It is, however, certainly true that once one of the modes grows into substantial amplitudes $\delta B/B_0 \sim 1$, other modes will also be strongly affected.

Another possible mechanism leading to saturation of the Weibel instability is the so-called \Alfven current limit \citep{Kato2005,Achterberg2007b}. Physically, this limit may be understood as the condition in which the cylindrical electric current produced by a particle stream generates a magnetic field that is so strong that the particle stream itself cannot be maintained due to the magnetic deflection. It is roughly the same as the condition that the radius of the cylindrical current becomes comparable to the gyroradius of particles defined with the magnetic field generated by the current.

We here consider a slightly modified idea by replacing the comparison of spatial scales with time scales \citep{Bret2013,Ruyer2015}. Namely, one may conjecture that the instability may be suppressed when the growth rate becomes comparable to the gyrofrequency defined by the wave magnetic field:
\begin{align}
    \gamma(k) \sim \frac{e B_k}{m_i c}.
    \label{eq:alfven-cond}
\end{align}
If we substitute Eq.~(\ref{eq:cold-weibel}) to the above estimate, we obtain $\varepsilon_{B} \sim (n_r/2 n_0) / (1 + \wpe^2/(k^2 c^2))$. We confirm that, for $k c/\wpe \gg 1$, it is equivalent to the original \Alfven current limit. The screening effect for $k c/\wpe \lesssim 1$ introduces finite deviation, because only a fraction of the ion current can contribute to the generation of the magnetic field. The saturation level in this scenario for an arbitrary growth rate may then becomes:
\begin{align}
    \varepsilon_{B} \sim
    \left( \frac{\gamma(k)}{\Wci} \right)^2
    \MA^{-2}
    \label{eq:epsilon-alfven}
\end{align}
which we call the \Alfven current limit. Although we here define the saturation level formally as a function of wavenumber for comparison presented below, the actual saturation may be determined by the wave power integrated over wavenumber space.

We note that similar estimates for the saturation level and comparisons with nonlinear simulations have already been given previously for a slightly different context. Perhaps, it is interesting to mention that the trapping and \Alfven current limits both give the same order of magnitude estimates in relativistic pair plasmas \citep{Bret2013}. As we will see below, in general, they may predict substantially different saturation levels depending on the growth rate and the dominant wavenumber.

\begin{figure}[t]
    \centering
    \includegraphics[height=0.7\textheight,width=0.5\textwidth,keepaspectratio]{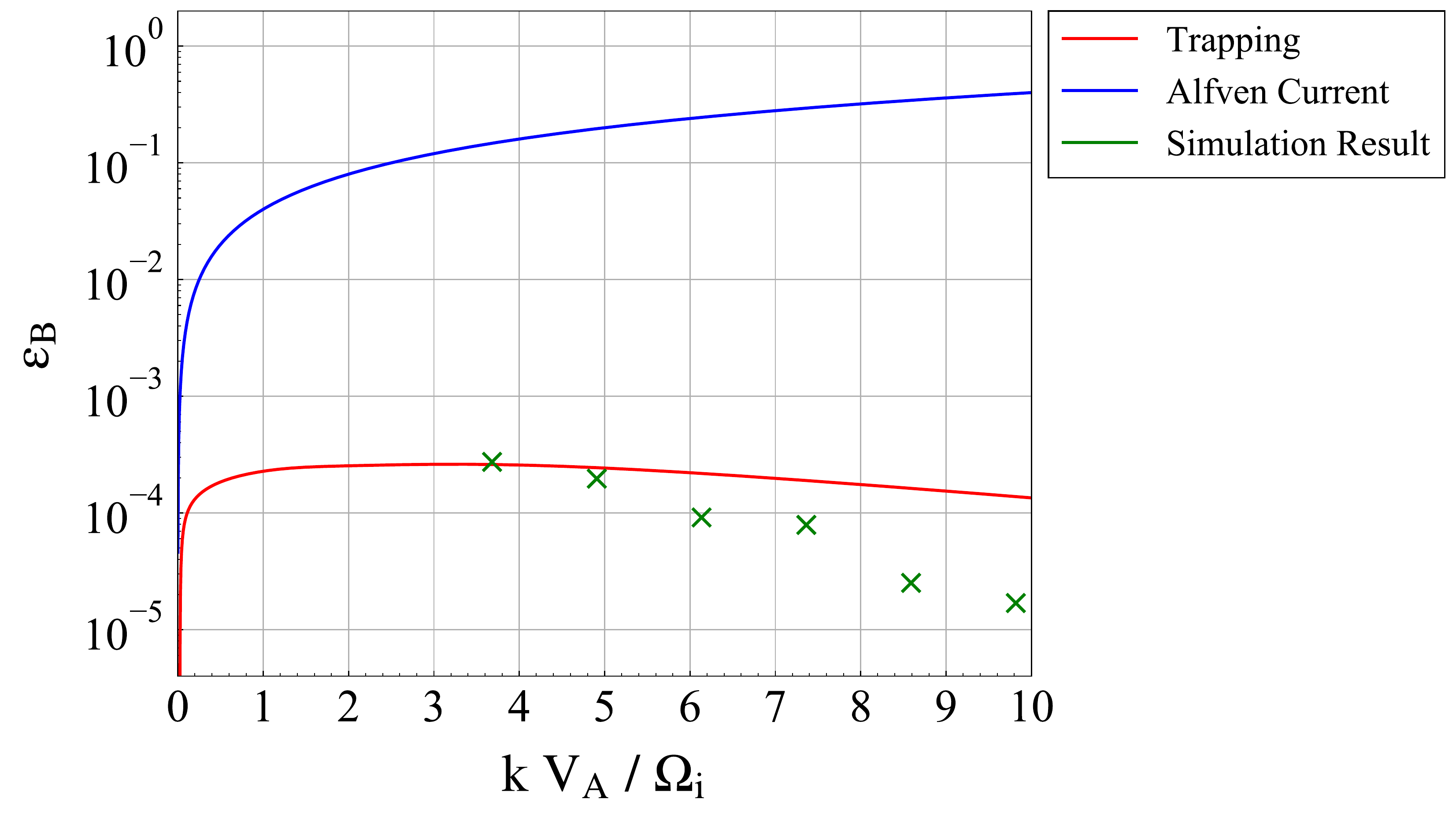}
    \caption{
        Comparison of saturation levels between theory and simulation. The saturation levels evaluated for each mode from Run 1 are shown with crosses. Theoretical saturation levels are shown with solid lines for the trapping limit in red and the \Alfven current limit in blue.
    }
    \label{fig:saturation-run1}
\end{figure}
   
\begin{figure}[t]
    \centering
    \includegraphics[height=0.7\textheight,width=0.5\textwidth,keepaspectratio]{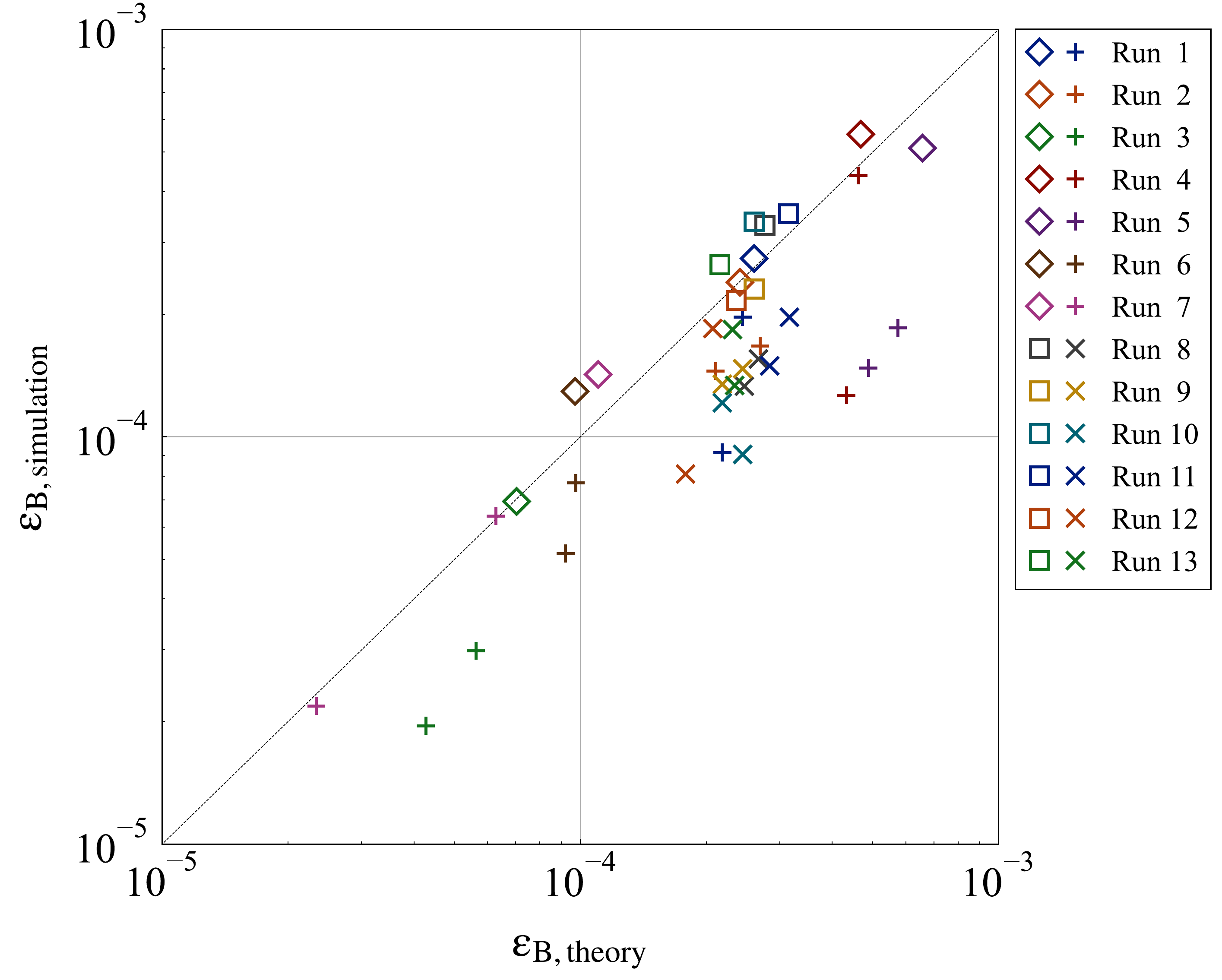}
    \caption{
        Compilation of saturation levels evaluated from simulations as compared with theory. The vertical axis indicates the saturation levels from the simulations, whereas the horizontal axis indicates the corresponding theoretical estimates based on the trapping limit. The results from different runs are represented by different colors. The three largest amplitude modes at saturation are selected from each simulation: The diamonds (Run 1 to 7) and squares (Run 8 to 13) indicate the largest amplitude mode, while the pluses (Run 1 to 7) and crosses (Run 8 to 13) indicate the remaining two modes.
    }
    \label{fig:saturation-summary}
\end{figure}

Fig.~\ref{fig:saturation-run1} compares the simulation results obtained for Run 1 and the above two theoretical estimates. The theoretical estimates are calculated by substituting the growth rate $\gamma(k)$ obtained from the fully kinetic dispersion relation to Eqs.~(\ref{eq:epsilon-trapping}) and (\ref{eq:epsilon-alfven}). We evaluate the saturation levels of the simulation from the peak amplitude immediately after the clear linear growth phase. (We ignore some modes for which a peak immediately after the linear growth phase could not be clearly identified.) We see that the trapping limit predicts a much smaller saturation level than the \Alfven current limit because Eq.~(\ref{eq:epsilon-trapping}) has a much stronger dependence on $\MA$ than Eq.~(\ref{eq:epsilon-alfven}). The simulation results shown with crosses appear mostly around the trapping limit estimate or below. The \Alfven current limit gives clearly an overestimate by a few orders of magnitude, indicating that the saturation is likely to be determined by the particle trapping.

The reason why only a few of the modes can reach the theoretical estimate based on the particle trapping may be understood as follows. First, the estimate uses the growth rate calculated with the initial condition. As we have already seen, short-wavelength modes are subject to strong kinetic damping by an increased thermal velocity as a result of the quasi-linear heating in the early phase. Therefore, the saturation level must also be substantially lowered at short wavelengths according to Eq.~(\ref{eq:epsilon-trapping}). Second, the final saturation amplitude is of the same order of the background magnetic field $\delta B/B_0 \sim O(1)$. This indicates that, if one of the modes grows to the saturation amplitude, its effect on the other modes is no longer negligible. We have confirmed that only one or two modes reached the theoretical saturation level in the other simulation runs as well.

The saturation levels obtained with all the simulations in Table \ref{tab:parameters} are compiled in Fig.~\ref{fig:saturation-summary}, which are compared with the theoretical estimate based on the trapping limit. For this figure, we have selected the three largest amplitude modes at saturation from each simulation. For each mode, the saturation levels evaluated from the simulations and the corresponding theoretical estimates are shown in vertical and horizontal axes, respectively. The largest amplitude modes for each run are denoted by diamonds (Run 1 to 7) and squares (Run 8 to 13), whereas others are denoted by pluses (Run 1 to 7) and crosses (Run 8 to 13). It is clear that the theory and simulations are very well correlated. Because of the different dependence on $\gamma(k)$ and $\MA$, the \Alfven current limit always gives an estimate orders of magnitude larger than the trapping limit. Therefore, we think that the trapping limit should be a reasonable estimate for the saturation of the linear Weibel instability driven by the ion ring distribution.

Our simulation results are fully consistent with previous studies of the Weibel instability in an unmagnetized plasma \citep{Davidson1972,Ruyer2016}. The linear instability quenches by the particle trapping because the predicted saturation level is lower than the \Alfven current limit. In an unmagnetized plasma, current filaments generated during the linear phase start to merge with each other in the long-term evolution \citep{Medvedev2005,Achterberg2007b,Ruyer2016}. The magnetic field amplification associated with filament merging may proceed until the current reaches the critical \Alfven current. On the other hand, the nonlinear amplification in our simulations appears rather less efficient. The maximum amplitude typically reaches $\delta B/B_0 \sim O(1)$, which is of the same order as the saturation level predicted by the trapping limit, indicating that the filament merging is not as efficient as in an unmagnetized plasma.

The small box size used in our simulations is obviously one of the reasons why filament merging is prevented from generating large-scale structures with larger amplitudes. Another reason seems to be related to the magnetization of the driving particle population. Since the ions start to gyrate around the background magnetic field at $\Wci \, t \gtrsim 1$, it is natural that the development in this phase differs substantially from the unmagnetized plasma case. Indeed, we observe temporal oscillations of $B_x$ and $B_z$ in the late phase $\Wci \, t \gtrsim 1$, indicating that the fluctuations at this stage are Alfv\'enic in nature. In our typical simulation parameters, the time scale at which the linear instability reaches saturation is comparable to $\Wci^{-1}$. Therefore, nonlinear filament merging may be suppressed, or at least somehow modified, by the presence of a finite background magnetic field.

\subsubsection{Magnetic Reconnection}
\label{sec:reconnection}

\begin{figure}[t]
    \centering
    \includegraphics[height=0.7\textheight,width=0.5\textwidth,keepaspectratio]{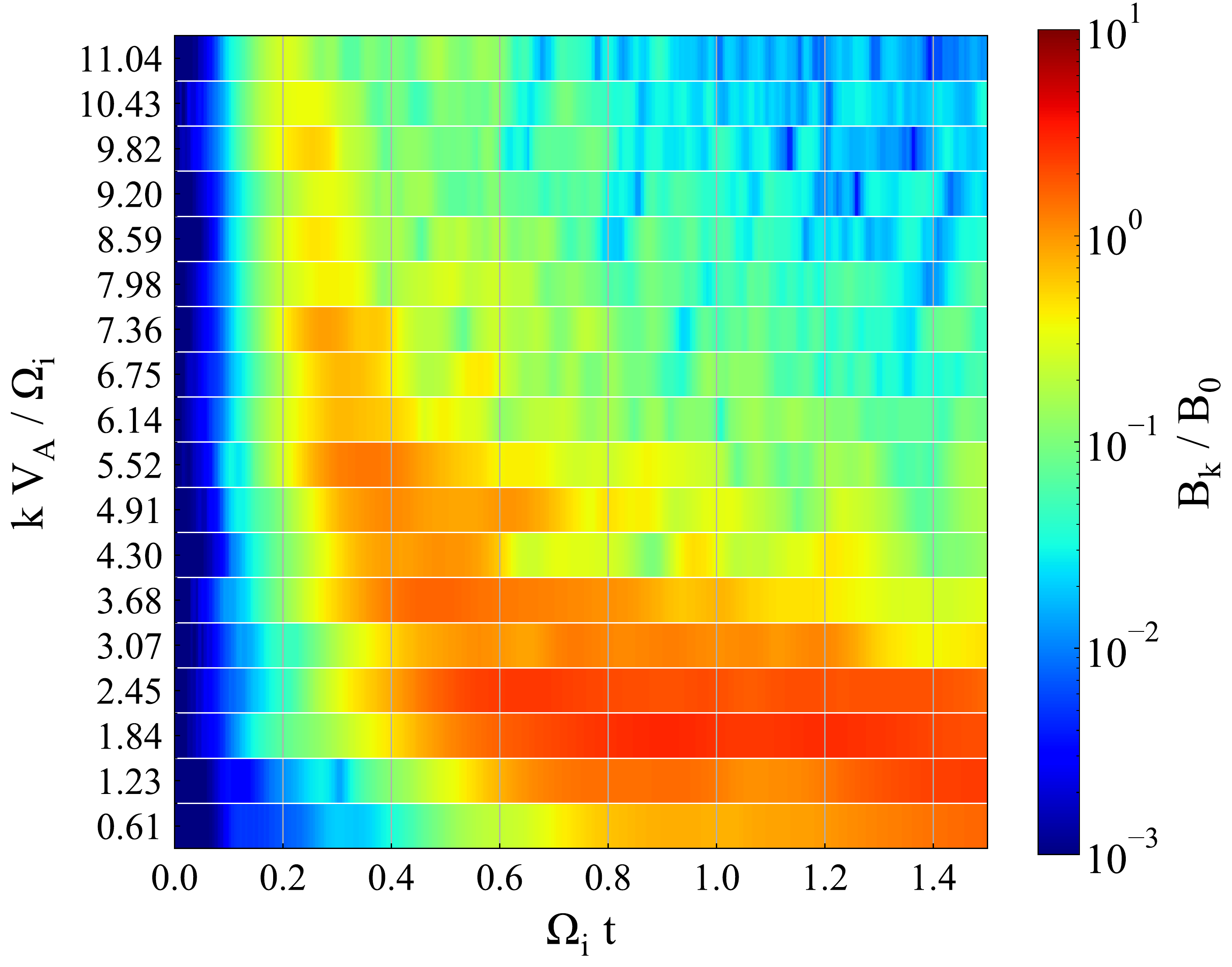}
    \caption{Time evolution of wavenumber spectrum for Run 13$^{\#}$. The format is the same as Fig.~\ref{fig:ky-spectrum}.}
    \label{fig:ky-reconnection}
\end{figure}

\begin{figure}[t]
    \centering
    \includegraphics[height=0.9\textheight,width=0.5\textwidth,keepaspectratio]{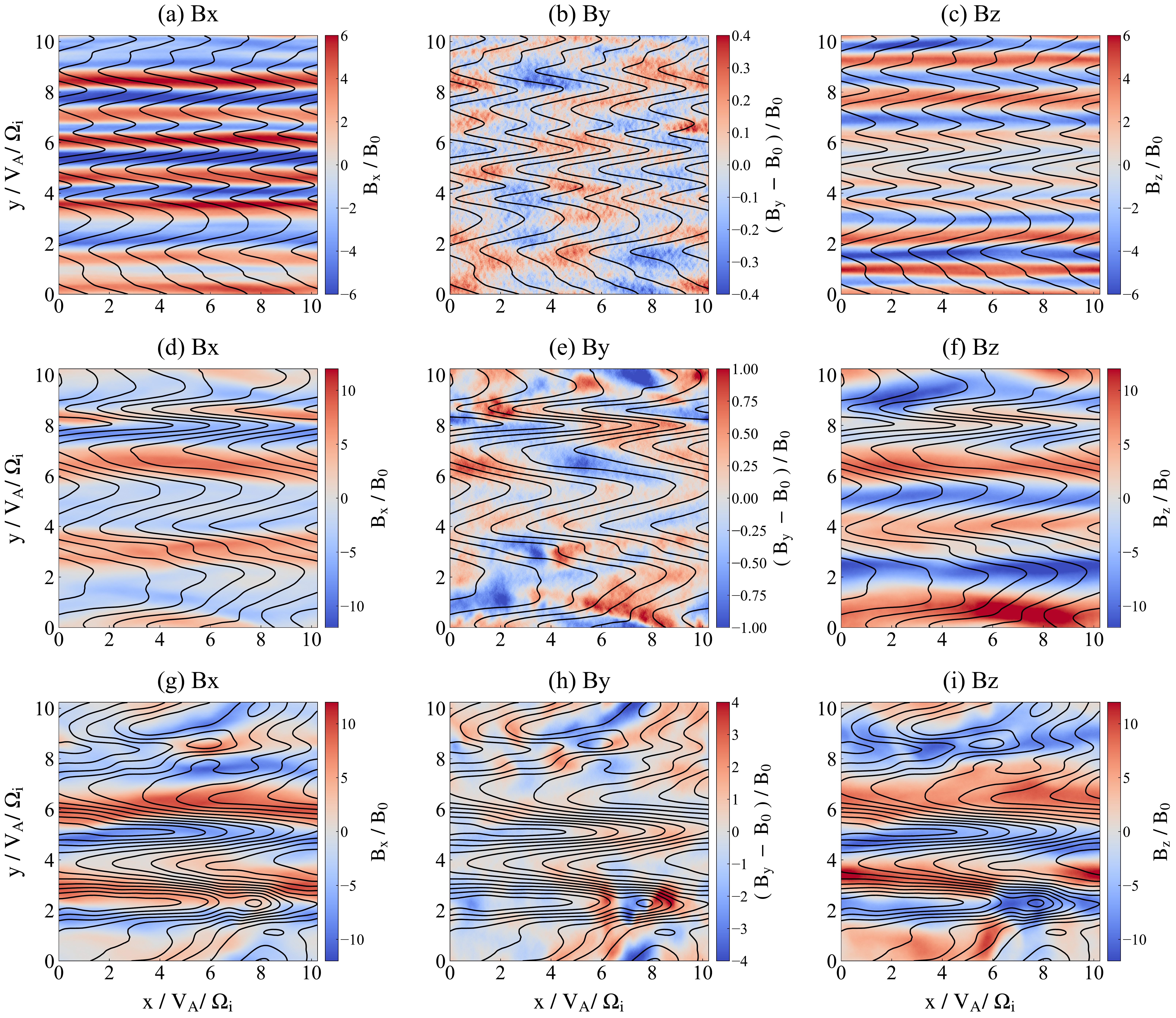}
    \caption{Snapshots of magnetic field for Run 13$^{\#}$. The format is the same as Fig.~\ref{fig:snapshots-bfield}. Top (a-c), middle (d-f), bottom (g-i) panels indicate snapshots taken at $\Wci \, t = 0.4, 0.8, 1.2$, respectively.}
    \label{fig:snapshot-reconnection}
\end{figure}

The agreement between the simulations and the theoretical estimate based on the trapping limit might be seen as a disappointment because it gives the saturation level much lower than the \Alfven current limit. A typical saturation level of $\delta B/B_0 \sim O(1)$ predicted by the trapping limit does not appear to be sufficient to drive magnetic reconnection in the nonlinear phase. On the other hand, magnetic reconnection has been observed in PIC simulations that model the global structure of strong perpendicular shocks. This indicates that the present model might not be adequate in describing the fully nonlinear dynamics of the shock quantitatively. Nevertheless, we here demonstrate that magnetic reconnection is indeed triggered if a certain condition is satisfied.

Since the trapping-limited saturation amplitude of the linear instability is not sufficient to produce $\delta B/B_0 \gg 1$, the nonlinear filament merging is needed for further amplification. This may happen if the growth rate is so large that the nonlinear phase begins before the reflected ions become magnetized. In addition, the simulation box size needs to be sufficiently large for the production of large-scale structures. We conducted simulations with larger box sizes than our fiducial choice of $5 \, \VA/\Wci$ and found signatures of magnetic reconnection in a run with the same parameters as Run 13 when the box size was doubled. In the following, we will discuss this run, which will be referred to as Run 13$^{\#}$. Note that the linear growth rate of Run 13 is the largest of all in Table.~\ref{tab:parameters} because of the large reflected ion density $n_r/n_0 = 0.5$.

Fig.~\ref{fig:ky-reconnection} shows the time evolution of the parallel wavenumber spectrum. It is seen that higher wavenumber modes $k \VA/\Wci \gtrsim 4$ reach the saturation relatively earlier times $\Wci \, t \sim 0.3$. The lower wavenumber modes grow later to reach larger amplitudes. In this particular case, the mode $k \VA/\Wci = 1.84$ (mode 3) has reached well beyond the saturation level predicted by the trapping limit.

The snapshots of the magnetic field at $\Wci \, t = 0.4, 0.8, 1.2$ are shown in Fig.~\ref{fig:snapshot-reconnection}. We again confirm that the instability generates perturbations primarily in $B_x$ and $B_z$ in the early phase $\Wci t = 0.4$. As the system evolves in time, larger-scale folded current sheets are generated, and finally, magnetic reconnection is triggered that produces multiple magnetic islands. The substantial perturbations in $B_y$ at $\Wci \, t = 1.2$ may be attributed to the result of magnetic reconnection because it will naturally produce $B_y$ (i.e., the reconnected component) from the $x$ component of the magnetic field generated by the instability. Visual inspection of the magnetic field line motion also confirms that magnetic reconnection is actually occurring in the late nonlinear phase.

The purpose of this demonstration is to show that magnetic reconnection may indeed occur in our model when the instability is driven strongly enough in a large system. However, we think that it is difficult to estimate quantitatively the condition that is required for such spontaneous magnetic reconnection triggering with the simplified model. As we will see below in Section~\ref{sec:discussion}, the condition for the instability to behave Weibel-like ($\gamma/\Wci \gg 1$) itself makes the gyrotropic assumption for the reflected ions rather questionable.

\section{Discussion and Conclusions}
\label{sec:discussion}
In this paper, we have investigated the ion-scale kinetic instability driven by the reflected ions in the transition layer of collisionless strictly perpendicular shocks. For this purpose, we have adopted a homogeneous plasma model with the reflected ion distribution function represented by a ring distribution in velocity space. Linear dispersion analysis for parallel propagating waves is performed, which finds that the AIC and Weibel instabilities are both on the same branch and become dominant in different limiting cases. In terms of macroscopic shock parameters, the instability characteristics are largely controlled by both the \Alfven and sound Mach numbers. Both Mach numbers should be sufficiently large for the instability to appear as Weibel-like. Otherwise, an AIC-like instability will result. We suggest that these two regimes may be distinguished in terms of $\gamma/\Wci$. For $\gamma/\Wci \lesssim 1$, the reflected ions are magnetized, and we may obtain the classical AIC instability. In the opposite case $\gamma/\Wci \gg 1$, they behave essentially unmagnetized, which is consistent with the fact that the instability has similar characteristics to the Weibel instability in an unmagnetized plasma.

Previous works \citep{Davidson1972,Achterberg2007a} on the Weibel instability adopting an unmagnetized plasma model for linear analysis have shown that the growth rate of the instability is proportional to $V_s/c$. However, if the growth rate is normalized to the ion gyrofrequency, it is proportional to the Alfvén Mach number in the cold plasma limit. The physical insight from this change of normalization had not been previously appreciated. Qualitatively, a finite magnetic field effect eliminates the electron screening effect at long wavelengths and thus enhances the growth rate. Our finding suggests that young SNR shocks with $V_s/c \sim O(0.01)$ are likely to be Weibel-dominated shocks even though the shock speed is much lower than the speed of light as well as typical shock speeds used routinely for PIC simulations of non-relativistic magnetized shocks $V_s/c \sim O(0.1)$. Furthermore, planetary bow shocks in the heliosphere can potentially be in this regime if both the solar-wind magnetic field strength and temperature are unusually small. Although high \Alfven Mach number shocks are occasionally observed at the terrestrial bow shock \citep{Sundberg2017a,Madanian2020}, we suggest that relatively high plasma betas (i.e., lower sound Mach numbers) at these shock will likely make the rippling, rather than the Weibel, as the dominant mode. The situation might change at outer solar system planets because adiabatic cooling of the solar wind plasma will naturally lower the temperature at large heliocentric distances.

We have also performed 2D PIC simulations with the homogeneous model, focusing in particular on the parameter regime where the Weibel instability becomes dominant. Although the linear analysis is limited for modes propagating along the ambient magnetic field, we see no evidence for significant oblique modes in the simulations. Therefore, the obliquely propagating mirror instability has been ruled out at high Mach numbers. The electron-scale instabilities that grow and saturate much more rapidly do not affect the ion-scale instability. We find that the saturation level of the linear Weibel instability is consistent with a theoretical model based on the particle trapping by the wave magnetic field. Subsequent magnetic field amplification is rather inefficient as compared to the unmagnetized plasma case, probably because the nonlinear filament merging is affected by the presence of a finite background magnetic field. Consequently, the maximum magnetic field amplitude is typically of the same order of the background magnetic field $\delta B/B_0 \sim O(1)$. We have shown that, if an extreme condition is considered, $\delta B/B_0 \gg 1$ is reached, and magnetic reconnection occurs spontaneously. Formally, this implies that spontaneous magnetic reconnection suggested previously is unlikely to occur unless the \Alfven Mach number of the shock is extremely large, larger than what is expected for a typical young SNR shock. As we see below, however, the present paper is not able to draw quantitative conclusions on this issue.

Our intention to adopt the gyrotropic velocity distribution for the reflected ions is to discuss both the AIC and Weibel instabilities on equal footing. The AIC instability originating from the cyclotron resonance should assume a gyrotropic distribution in its equilibrium state. On the other hand, the Weibel instability may be driven by an unmagnetized or non-gyrotropic ion beam. Recall that the gyrotropic ring distribution is essentially a velocity distribution that results from spatial integration over the shock transition layer. The large growth rate $\gamma/\Wci \gg 1$ implies that the typical scale length over which the perturbation grows into large amplitudes $V_s/\gamma$ is much smaller than the thickness of the shock. Therefore, the reflected ions are likely to appear as a non-gyrotropic beam in the wave growth region. We note that the instability does indeed grow rapidly in the leading edge of the shock in published 2D PIC simulation results that model the entire shock structure \citep{Matsumoto2015,Bohdan2020a}. In other words, while the gyrotropic distribution is reasonable to differentiate the instability property, it may not necessarily be a quantitatively appropriate model to investigate the nonlinear dynamics in the Weibel-dominated shock.

Preliminary PIC simulations of the instability driven by unmagnetized ion beams with a homogeneous model have shown that the growth rate is typically larger than that presented in this paper. Furthermore, magnetic reconnection seems to occur more easily in the nonlinear phase at \Alfven Mach numbers lower than with the gyrotropic ring distribution model. Perhaps, a larger growth rate driven by the unmagnetized non-gyrotropic beam will push the system into the filament merging phase before the ions behave magnetized $\Wci \, t \ll 1$. The nonlinear evolution then becomes essentially the same as the unmagnetized plasma case, in which the magnetic field may be amplified to reach essentially the \Alfven current limit \citep{Kato2010,Matsumoto2015,Bohdan2021}. On the other hand, the saturation in most of our simulations presented in this paper occurs typically at $\Wci \, t \sim 1$, beyond which the ions start to behave magnetized. Although a very large temperature anisotropy $T_{\perp}/T_{\para} \gg 1$ has yet remained in the saturation phase, an increased parallel temperature of the reflected ions will make the further development of the system AIC-like. This may be the reason why we observe the amplification and subsequent evolution are less violent in the gyrotropic ring distribution model.

Another important aspect is that the instability in an actual shock is continuously driven. The free energy in a homogeneous model is given only at the initial condition and will be transferred to other forms of energy as a result of the instability. In contrast, the free energy at a realistic shock is flowing into the system at a constant rate from the upstream. This may introduce substantial differences in the nonlinear outcome, as have been found previously for the high-frequency electrostatic Buneman instability \citep{Ohira2007,Amano2009a,Amano2009b}.

In a realistic Weibel-dominated shock, the non-gyrotropic unmagnetized reflected ion beam will drive the Weibel instability first in the leading edge of the shock front. The filament merging will then proceed in a region deeper in the transition layer to further amplify the magnetic field, possibly to a level predicted by the \Alfven current limit. At some point, the magnetic field has been strongly amplified $\delta B/B_0 \gg 1$, and magnetic reconnection may be initiated spontaneously. In the immediate downstream of the shock, the reflected ions become nearly gyrotropic with an increased thermal spread due to the heating in the transition layer. If the heating within the shock is not sufficient, the gyrotropic reflected ions may drive the AIC or mirror instabilities in the downstream region for further, but slower, relaxation of the anisotropy. Therefore, we suggest that detailed nonlinear evolution of the Weibel instability and subsequent dynamics in a strong shock should be studied with a non-gyrotropic reflected ion distribution, preferably with an open-boundary model. Such investigation will be crucial to draw a quantitative conclusion on the condition required for magnetic reconnection and to quantify its role in the heating and acceleration of particles.

Finally, we suggest that the transition of the characteristics between different regimes may have an important impact on particle acceleration. The plasma wave activities within the shock transition layer significantly affect the efficiency of particle acceleration, in particular, the injection of electrons into the first-order Fermi acceleration process. We have recently proposed the stochastic shock drift acceleration (SSDA) as a plausible electron injection mechanism that is fully consistent with in-situ observations at the terrestrial bow shock \citep{Katou2019,Amano2020}. We have also seen a signature of SSDA in a fully three-dimensional Weibel-dominated shock \citep{Matsumoto2017}. Nevertheless, the magnetic reconnection activity appeared rather low in 3D compared to 2D results \citep{Matsumoto2015,Bohdan2020a}, which might be due to reduced parameters adopted in 3D. It should be interesting to investigate if SSDA survives in a very high Mach number SNR shock dominated by the Weibel instability undergoing violent magnetic reconnection.

\begin{acknowledgments}
This work was supported by JSPS KAKENHI grant Nos.~17H02966 and 17H06140. This work used the computational resources of the HPCI system provided by the Information Technology Center, Nagoya University, through the HPCI System Research Project (Project ID: hp200035). Numerical simulations were also carried out with Cray XC50 at the Center for Computational Astrophysics, National Astronomical Observatory of Japan,  the KDK computer system at Research Institute for Sustainable Humanosphere, Kyoto University.
\end{acknowledgments}

\section*{data availability}
The data that support the findings of this study are available from the corresponding author
upon reasonable request.

\appendix
\section{Dependence on Other Parameters}
\label{sec:appendix}
Here we show explicitly the dependence of the instability on the following parameters: $n_r/n_0$, $\beta_e$, $m_i/m_e$, and $\wpe/\Wce$.

As has become clear with the approximate analytic expressions obtained with the cold plasma approximation, the growth rate in unit of the ion gyrofrequency ($\gamma/\Wci$) is controlled primarily by the reflected ion density $n_r/n_0$ and the \Alfven Mach number $\MA$. The increase in the reflected ion density and the \Alfven Mach number has a qualitatively similar effect on the growth rate, which we have confirmed through an extensive parameter survey. Although a fixed reflected ion density of $n_r/n_0 = 0.2$ has been used throughout the linear analysis shown in the paper, this does not change the conclusion because the range of possible variation is quite limited, and the dependence can be easily understood.

\begin{figure}[tb]
    \centering
    \includegraphics[height=\textheight,width=0.5\textwidth,keepaspectratio]{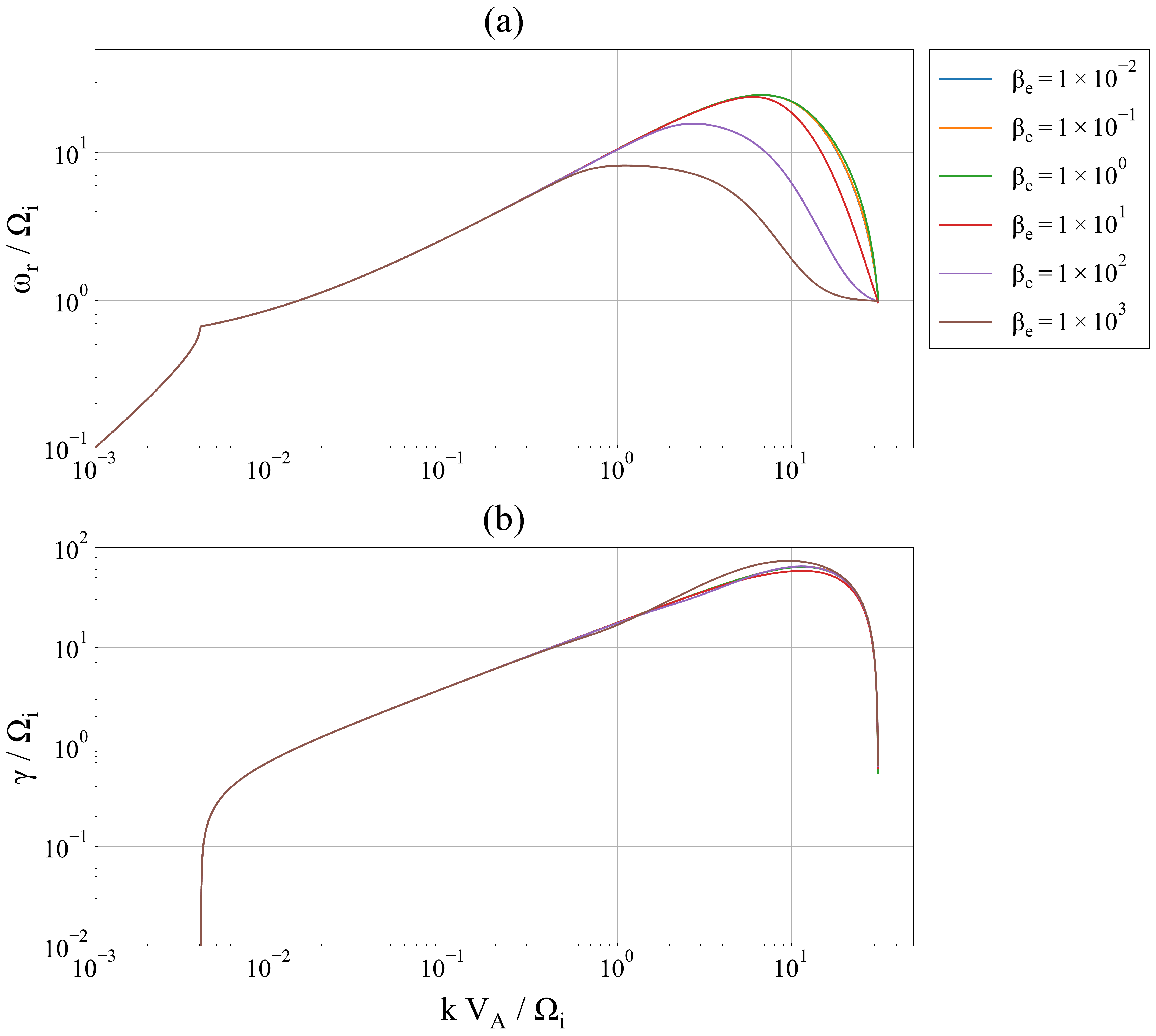}
    \caption{
        $\beta_e$ dependence of unstable solution. (a) real, and (b) imaginary parts of the numerical solution for $\beta_e = 1 \times 10^{-2}, 1 \times 10^{-1}, 1 \times 10^{0}, 1 \times 10^{1}, 1 \times 10^{2}, 1 \times 10^{3}$ are shown. The other parameters are: $n_r/n_0 = 0.2$, $\MA = 300$, $\MS = 100$, $m_i/m_e = 1836$, $\Wce / \wpe = 5 \times 10^{-3}$.
    }
    \label{fig:betae-dep}
\end{figure}

\begin{figure}[tb]
    \centering
    \includegraphics[height=\textheight,width=0.5\textwidth,keepaspectratio]{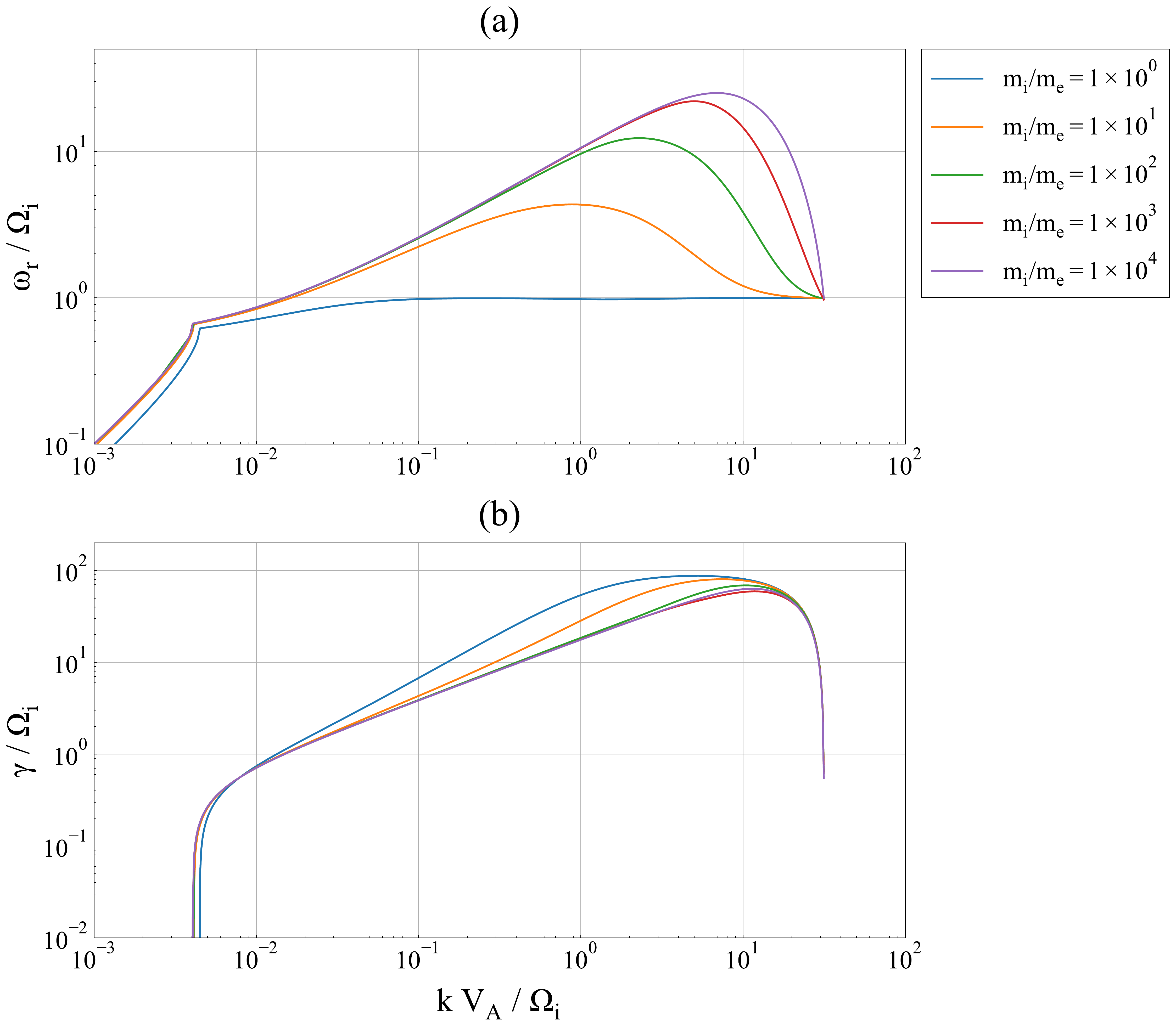}
    \caption{
        $m_i/m_e$ dependence of unstable solution. (a) real, and (b) imaginary parts of the numerical solution for $m_i/m_e = 1 \times 10^{0}, 1 \times 10^{1}, 1 \times 10^{2}, 1 \times 10^{3}, 1 \times 10^{4}$ are shown. The other parameters are: $n_r/n_0 = 0.2$, $\MA = 300$, $\MS = 100$, $\Wce / \wpe = 5 \times 10^{-3}$, $\beta_e =10$.
    }
    \label{fig:mime-dep}
\end{figure}

We have found that the growth rate in the Weibel regime $\gamma/\Wci \gg 1$ depends only weakly on $\beta_e$ and $m_i/m_e$. Figs.~\ref{fig:betae-dep} and \ref{fig:mime-dep} respectively show the dependence on $\beta_e$ and $m_i/m_e$. We use $m_i/m_e = 1836$ for Fig.~\ref{fig:betae-dep} and $\beta_e = 10$ for Fig.~\ref{fig:mime-dep}. The other parameters are fixed: $n_r/n_0 = 0.2$, $\MA = 300$, $\MS = 100$, $\Wce / \wpe = 5 \times 10^{-3}$. In both cases, decreasing $m_i/m_e$ or increasing $\beta_e$ reduces the real frequency at the short wavelength regime near the peak growth rate, whereas the maximum growth rate itself is only weakly modified. We note that, since we normalize the time and space with ion scale quantities, changing the mass ratio $m_i/m_e$ indicates changing the electron response, in particular the thermal velocity. We may expect that the electron kinetic effect becomes important when the thermal electrons satisfy the cyclotron resonance condition. The parameter of interest $\omega/\Wci \ll m_i/m_e$ indicates that the resonance condition may be written by $k \vth{e} \sim |\Wce|$, or equivalently, $k \VA/\Wci \sim ( m_i/m_e )^{1/2} \beta_e^{-1/2}$. At this wavenumber, the electrons are no longer strictly magnetized because the electron thermal gyroradius is of the same order of the wavelength of fluctuations. Therefore, the electrons behave essentially incoherent background particles that do not respond to the perturbations. In contrast to the similar estimate for the reflected ions Eq.~(\ref{eq:critical-k}), the driving force of the instability itself is little modified by the unmagnetized electrons. This gives a reasonable interpretation for the numerical solutions in which the imaginary part is not sensitive to the electron response. Although the maximum growth shifts to a smaller wavenumber at a very small mass ratio $m_i/m_e \lesssim 10$, a reduced mass ratio of $m_i/m_e \gtrsim 100$ will be sufficient to reproduce a correct linear phase in PIC simulations.

We have also investigated the dependence on the magnetization $\Wce/\wpe$, in which no appreciable change is found both for the real and imaginary part of the unstable solution in the range $5 \times 10^{-4} \leq \Wce/\wpe \leq 5 \times 10^{-1}$. This result is consistent with the prediction by the cold plasma approximation, in which the unstable solution does not depend on the magnetization. The kinetic effect does not introduce any differences in this respect. We have confirmed that the unstable fully kinetic solution always satisfies the condition $\omega/kc \ll 1$, in which case the displacement current can be ignored for low-frequency modes, and the speed of light should not appear explicitly in the dispersion relation. Since the magnetization is a parameter associated with the \Alfven speed relative to the speed of light $\VA/c$ (as mentioned in Section \ref{sec:model}), it is natural that a low-frequency instability is little affected by the magnetization if appropriately normalized.



\bibliographystyle{plain}
%


\end{document}